\pdfoutput=1 

\documentclass[useAMS,usenatbib]{mnras}
\usepackage{graphicx, color}
\usepackage{xspace}
\usepackage{epsfig}
\usepackage{rotate}
\usepackage{latexsym}
\usepackage{amssymb}
\usepackage{psfrag}
\usepackage{lineno}
\usepackage{amsmath}
\usepackage{ae,aecompl}

\def\aap{{ A\&A}}

\def\aj{{AJ}}

\def\apj{{ApJ}}

\def\mnras{{MNRAS}}
\def\prd{{PRD}}

\def\jcap{JCAP}
\def\pasj{{PASJ}}
\def\nat{Nature}

\def\apjs{{ApJS}}

\def\vk{{\hbox{\bf k}}}
\def\vq{{\hbox{\bf q}}}

\def\Mpc{\, h^{-1} \, {\rm Mpc}}

\def\kvecMpc{\, h \, {\rm Mpc}^{-1}}
\def\kvecMpccube{\, h^3 \, {\rm Mpc}^{-3}}
\def\Msun{\, h^{-1} \, {\rm M}_\odot}

\def\halogen{{\sc{Halogen}}\xspace}
\def\icecola{{\sc{ICE-COLA}}\xspace}
\def\patchy{{\sc{Patchy}}\xspace}
\def\pinocchio{{\sc{Pinocchio}}\xspace}
\def\peakpatch{{\sc{PeakPatch}}\xspace}

\newcommand{\beq}{\begin{equation}}
\newcommand{\eeq}{\end{equation}}
\newcommand{\beqa}{\begin{eqnarray}}
\newcommand{\eeqa}{\end{eqnarray}}

\begin{document}

\title[Power spectrum covariance comparison]{
Comparing approximate methods for mock catalogues and covariance matrices II: Power spectrum multipoles}

\author[L. Blot et al.]{Linda Blot$^{1,2,\star}$ %\thanks{E-mail:blot@ice.cat},
Martin Crocce$^{1,2}$,
Emiliano Sefusatti$^{3,4}$,
Martha Lippich$^{5,6}$,
\newauthor
Ariel G. S\'{a}nchez$^{5}$,
Manuel Colavincenzo$^{7,8,9}$,
Pierluigi Monaco$^{9,3,4}$,
\newauthor
Marcelo A. Alvarez$^{10}$,
Aniket Agrawal$^{11}$,
Santiago Avila$^{12}$,
\newauthor
Andr\'es Balaguera-Antol\'{i}nez$^{13,14}$,
Richard Bond$^{15}$,
Sandrine Codis$^{15,16}$,
\newauthor
Claudio Dalla Vecchia$^{13,14}$,
Antonio Dorta$^{13,14}$,
Pablo Fosalba$^{1,2}$,
\newauthor
Albert Izard$^{17,18,1,2}$,
Francisco-Shu Kitaura$^{13,14}$,
Marcos Pellejero-Ibanez$^{13,14}$,
\newauthor
George Stein$^{15}$,
Mohammadjavad Vakili$^{19}$,
Gustavo Yepes$^{20,21}$.
\\ \\
% List of institutions
$^{1}$ Institute of Space Sciences (ICE, CSIC), Campus UAB, Carrer de Can Magrans, s/n,  08193 Barcelona, Spain \\
$^{2}$ Institut d'Estudis Espacials de Catalunya (IEEC), 08034 Barcelona, Spain \\
$^{3}$ Istituto  Nazionale  di  Astrofisica, Osservatorio Astronomico di Trieste, via Tiepolo 11, 34143 Trieste, Italy\\
$^{4}$ Istituto  Nazionale  di  Fisica  Nucleare, Sezione di Trieste, Via Valerio, 2, 34127 Trieste, Italy\\
$^{5}$ Max-Planck-Institut f\"ur extraterrestrische Physik, Postfach 1312, Giessenbachstr., 85741 Garching, Germany\\
$^{6}$ Universit\"ats-Sternwarte M\"unchen, Ludwig-Maximilians-Universit\"at M\"unchen, Scheinerstrasse 1, 81679 Munich, Germany\\
$^{7}$ Dipartimento  di  Fisica,  Universit\`a  di  Torino,  Via  P.  Giuria  1,  10125  Torino,  Italy\\
$^{8}$ Istituto  Nazionale  di  Fisica  Nucleare,  Sezione  di  Torino,  Via  P.  Giuria  1,  10125  Torino,  Italy\\
$^{9}$ Dipartimento di Fisica, Sezione di Astronomia, Universit\`a di Trieste, via Tiepolo 11, 34143 Trieste, Italy\\
$^{10}$ Berkeley Center for Cosmological Physics, Campbell Hall 341, University of California, Berkeley CA 94720, USA \\
$^{11}$ Max-Planck-Institut f\"ur Astrophysik, Karl-Schwarzschild-Str. 1, 85741 Garching, Germany\\
$^{12}$ Institute of Cosmology \& Gravitation, Dennis Sciama Building, University of Portsmouth, Portsmouth PO1 3FX, UK\\
$^{13}$ Instituto de Astrof\'isica de Canarias, C/V\'ia  L\'actea, s/n, 38200, La Laguna, Tenerife, Spain\\
$^{14}$ Departamento Astrof\'isica, Universidad de La Laguna,  38206 La Laguna, Tenerife, Spain\\
$^{15}$ Canadian Institute for Theoretical Astrophysics, University of Toronto, 60 St. George Street, Toronto, ON M5S 3H8, Canada\\
%$^{16}$ CNRS \& Sorbonne Universit\'e, UMR 7095, Institut
%d'Astrophysique de Paris, 75014, Paris, France\\
$^{16}$ Institut d'Astrophysique de Paris, CNRS \& Sorbonne Universit\'e, UMR 7095, 98 bis boulevard Arago, 75014 Paris, France \\
$^{17}$ Jet Propulsion Laboratory, California Institute of Technology, 4800 Oak Grove Drive, Pasadena, CA 91109, USA\\
$^{18}$ Department of Physics and Astronomy, University of California, Riverside, CA 92521, USA\\
$^{19}$ Leiden Observatory, Leiden University, P.O. Box 9513, 2300 RA, Leiden, The Netherlands \\
$^{20}$ Departamento de F\'isica Te\'orica, M\'odulo 15, Universidad Aut\'onoma de Madrid, 28049 Madrid, Spain\\
$^{21}$ Centro de Investigaci\'on Avanzada en F\'isica Fundamental (CIAFF), Universidad Aut\'onoma de Madrid, 28049, Madrid, Spain\\
\vspace{-0.1cm}\\
$^\star${\tt E-mail:blot@ice.cat}
}

\maketitle
\label{firstpage}

\begin{abstract} 

We study the accuracy of several approximate methods for gravitational dynamics in terms of 
halo power spectrum multipoles and their estimated covariance matrix. We propagate the differences in covariances into parameter constrains related to growth rate of structure, Alcock-Paczynski distortions and biasing. We consider seven methods in three broad categories: algorithms that solve for halo density evolution deterministically using Lagrangian trajectories
(\icecola, \pinocchio and \peakpatch), methods that rely on halo assignment schemes onto dark-matter overdensities calibrated with a target N-body run (\halogen, \patchy) and two standard assumptions about the full density PDF (Gaussian and Lognormal). We benchmark their performance against a set of three hundred N-body simulations, running similar sets of approximate simulations with matched initial conditions, for each method. We find that most methods reproduce the monopole to within $5\%$, while residuals for the quadrupole are sometimes larger and scale dependent. The variance of the multipoles is typically reproduced within $10\%$. Overall, we find that covariances built from approximate simulations yield errors on model parameters within $10\%$ of those from the N-body based covariance. \end{abstract}


%%%%%%%%%%%%%%%%%%%%%%%%%%%%%%%%%%%%%%%%%%%%%%%%
% Introduction	     			       %
%%%%%%%%%%%%%%%%%%%%%%%%%%%%%%%%%%%%%%%%%%%%%%%%

\section{Introduction}
\label{sec:intro}
The study of the large-scale structure of the Universe has seen a major step up with the completion of the cosmological analysis of large galaxy redshift surveys such as SDSS \citep{2004PhRvD..69j3501T,2005ApJ...633..560E}, 2dFGRS \citep{PercivalEtal2001,2005MNRAS.362..505C} and more recently BOSS \citep{2017MNRAS.470.2617A} and VIPERS \citep{2017A&A...604A..33P}. We have entered an era in which the accuracy on cosmological parameters from the analysis of low redshift tracers becomes comparable to the one reached by Cosmic Microwave Background experiments, allowing tight constrains on parameters related to the late-time cosmic acceleration. This observational effort continues nowadays with the analysis of state-of-the-art imaging surveys for weak-lensing measurements: DES \citep{2017arXiv170801538T}, KiDS \citep{2017MNRAS.465.1454H} and HSC \citep{2018PASJ...70S..25M}, as well as campaigns reaching high redshift tracers like the Lyman-alpha forest and Quasars in eBOSS \citep{2016AJ....151...44D,2018MNRAS.473.4773A}. Even more excitingly,  the near future will see larger observational campaigns such as \textit{Euclid} \citep{LaureijsEtal2011}, DESI \citep{2016arXiv161100036D} or LSST \citep{2008arXiv0805.2366I,2009arXiv0912.0201L} that will enable a better understanding of several open questions, in addition to the nature of cosmic acceleration, such as the physics of the primordial universe and the neutrino mass scale. In parallel, the requirements on the accuracy in the evaluation of systematics and statistical errors have grown to become an important component of the error budget. In that sense, ensembles of galaxy mock catalogues are a vital component for the analysis of galaxy surveys, not only for their internal use but also for assessing the level of agreement among different datasets.
Mock catalogues are useful in at least three ways: (1) to study the impact of observational systematic effects, (2) to test science pipelines and analysis methodologies (including the recovery of the input cosmology of the mocks) and (3) to provide covariance matrices among observables (e.g. accounting for cosmic variance, noise, masking and correlated survey property fluctuations).

N-body simulations coupled with Semi-Analytical Models (SAM) or Halo Occupation Distribution (HOD) schemes for galaxy assignment are perhaps the best possible route nowadays to construct realistic galaxy catalogues \citep{2015MNRAS.447..646C,2017MNRAS.470.4646S,2017ApJ...850...24T}. However, running ensembles of high resolution N-body simulations of cosmological size is computationally very expensive \citep{2012MNRAS.426.2046A,2015MNRAS.448.2987F,2015ApJS..219...34H,2017ComAC...4....2P}, in particular for future surveys such as \textit{Euclid} or DESI, that will cover large volumes and will need high particle mass resolution to reach the expected galaxy number density in the observations.

One alternative route is to run fast algorithms (also known as ``approximate methods'', see \cite{monaco2016} for an overview) that reproduce, to a variable extent, the large-scale statistics of N-body simulations. This is done at the expense of losing accuracy in the small scale physics, since these methods are not able to resolve halo sub-structures. These approximate methods have been widely used over the past few years to help bring forward galaxy clustering analysis in a more comprehensive way \citep{Maneraetal2013,Maneraetal2015,Howlett2015,Kitaura2016,Koda2016,Avila2017}. Nonetheless an assessment of how those fast methods reproduce the covariance from N-body simulations, or their impact on derived cosmology, is still missing in the literature. Generally speaking, the requirements for future missions, such as {\it Euclid}, is that systematic errors in the estimation of covariance matrices should not bias the estimation of errors in cosmological parameters by more than $\sim 10\%$.

Hence, in this article we perform a robust and thorough comparison of clustering measurements in ensembles of mock catalogues produced from several state-of-the-art algorithms, that span basically all the various types of approaches available in the literature. We use as benchmark an ensemble of 300 large N-body simulations, short named Minerva \citep{GriebEtal2016}. As noted previously it is the first time this kind of study is performed. On the one hand because we concentrate on how well the observables are reproduced with high precision, thanks to the large number of realisations (see a previous comparison work by \cite{2015MNRAS.452..686C} using one N-body simulation). On the other hand because we extend the comparison to how well these methods reproduce the full covariance matrix of power spectrum multipoles, and how those inaccuracies propagate into constraints on the parameters of interest for galaxy redshift surveys. Two companion articles present similar studies in terms of two-point correlation function \citep{paper1} and bispectrum \citep{paper3}.

We span different types of approaches: one representing the combination of multi-step second order Lagrangian Perturbation Theory (LPT) to solve the large-scale displacements with a fast Particle Mesh (PM) solver to speed up the intermediate scale force computation, \icecola \citep{IzardCrocceFosalba2016,2015arXiv150904685I}. We then consider one group of methods based on halo finding in Lagrangian space coupled with LPT for the evolution equations of the collapsed regions. In this category we consider \pinocchio \citep{MonacoEtal2002} and \peakpatch \citep{BondMyers1996A,BondMyers1996B,BondMyers1996C,Steinetal2018}. Another group of methods uses LPT to evolve the matter field and then applies a biasing scheme to produce a halo sample. For this class we consider \halogen \citep{AvilaEtal2015} and \patchy \citep{Kitaura2014}. The latter methods need to be calibrated on one N-body simulation before running the desired set of mocks\footnote{Typically they require the halo abundance and bias as a function of  mass as measured in the parent simulation.}. Finally we consider approaches that make assumptions on the Probability Distribution Function (PDF) of the density field: a Gaussian model for the covariance \citep{GriebEtal2016} and a set of 1000 lognormal mocks \citep{AgrawalEtal2017}. Both of them use as input the actual clustering signal and number density of halos as measured in the benchmark N-body simulation. There are several other methods in the literature (e.g. {\sc QuickPM} by \cite{2014MNRAS.437.2594W}, {\sc FastPM} by \cite{2016MNRAS.463.2273F}, {\sc EZMocks} by \cite{2015MNRAS.446.2621C}) but they can all roughly fit into one of the above categories. Hence we expect our study to lead to conclusions that are of general applicability in the field.

This article is organised as follows: in Sec.~\ref{sec:methods} we describe all the approximate methods that we consider and how they compare in terms of computational cost. In Sec.~\ref{sec:samples} we introduce the halo samples over which we do the comparison. Sec.~\ref{sec:methodology} describes our methodology. Section~\ref{sec:results} contains and discusses our results, which are summarised in Sec.~\ref{sec:conclusions}. Lastly, we include two appendices, one describing an alternative way of defining the halo samples, and another discussing the statistical uncertainty in the comparison of cosmological parameter errors as shown in our results. 

%%%%%%%%%%%%%%%%%%%%%%%%%%%%%%%%%%%%%%%%%%%%%%%%
% Compared methods		               %
%%%%%%%%%%%%%%%%%%%%%%%%%%%%%%%%%%%%%%%%%%%%%%%%

\section{Compared methods}
\label{sec:methods}

We now briefly summarise the main features of the methods and mocks used in this work. For a comprehensive description we refer the reader to the first article of this series \citep{paper1}, while for a general review we refer to \cite{monaco2016}. For the purpose of presenting results, we will classify the compared methods in three categories: {\it predictive methods}, that do not require re-calibration against a parent N-body for each specific sample or cosmological model (\icecola, \pinocchio and \peakpatch), {\it calibrated methods}, that need prior information about the sample to simulate (\halogen and \patchy) and {\it analytical methods}, that predict the covariance by making assumptions on the shape of the PDF of the density field (lognormal and Gaussian). A summary is provided in Table 1, where we report the computing requirements (per single mock) for each of the methods used in this work, together with some general considerations and references. The methods are presented in decreasing computing time order.

Our comparison will focus on halo samples defined in comoving outputs at redshift $z=1$.

\begin{table*}
\begin{center}
\begin{tabular}{l l l l}
\hline
Method   &  Algorithm        &    Computational Requirements   &  Reference  \\ 
\hline   
Minerva     & {\bf N-body}        &  CPU Time: $4500$ hours                 &\cite{GriebEtal2016}  \\
         & Gadget-2  			&  Memory allocation: $660$ Gb
         & {\it https://wwwmpa.mpa-garching.mpg.de/} \\
         & Halos  : SubFind         & 		& {\it gadget/} \\ 
\hline
\hline   
\icecola     & Predictive        &  CPU Time:    $33$ hours              &\cite{IzardCrocceFosalba2016}  \\
         & 2LPT + PM solver  	&  Memory allocation: $340$ Gb   &
         Modified version of:  \\
         & Halos  : FoF(0.2)         &
         & {\it https://github.com/junkoda/cola\_halo}\\
\hline
\pinocchio     & Predictive        &  CPU Time:  $6.4$ hours                &\cite{MonacoEtal2013, MunariEtal2017}  \\
         & 3LPT + ellipsoidal collapse	 &  Memory allocation:  $265$ Gb
         & {\it https://github.com/pigimonaco/Pinocchio} \\
         & Halos  :  ellipsoidal collapse        &              & \\
\hline
\peakpatch     & Predictive        &  CPU Time:  $1.72$ hours$^*$                &\cite{BondMyers1996A, BondMyers1996B, BondMyers1996C}  \\
         & 2LPT + ellipsoidal collapse 			  &  Memory allocation:   $75$ Gb$^*$          & {\rm Not public}  \\
         & Halos  : Spherical patches                     &                                            &                   \\
         & over initial overdensities         &           & \\
\hline
\halogen     & Calibrated        &  CPU Time:  $0.6$ hours & \cite{AvilaEtal2015}.  \\
         & 2LPT + biasing scheme			  &  Memory allocation: $44$ Gb            & {\it https://github.com/savila/halogen}  \\
         & Halos  :  exponential bias        &   Input: $\bar{n}$, 2-pt correlation function,    & \\
         &	& halo masses and velocity field & \\
\hline
\patchy  & Calibrated        &  CPU Time:  $0.2$ hours                &\cite{Kitaura2014}  \\
         & ALPT + biasing scheme			  &  Memory allocation:  $15$
         Gb           & {\rm Not Public}  \\
         & Halos  : non-linear, stochastic        &    Input: $\bar{n}$, halo masses and       & \\
         & and scale-dependent bias 	& environment		& \\
\hline
Lognormal     & Calibrated        &  CPU Time:  $0.1$ hours                &\cite{AgrawalEtal2017}  \\
         & Lognormal density field 			  &  Memory
         allocation:  $5.6$ Gb           & {\it
           https://bitbucket.org/komatsu5147/}  \\       
         & Halos  :  Poisson sampled points        				&      Input:  $\bar{n}$, 2-pt correlation function     & {\it
           lognormal\_galaxies}  \\
\hline
Gaussian     & Theoretical       &  CPU Time:  n/a                &\cite{GriebEtal2016} \\
         & Gaussian density field 			  &  Memory allocation:  n/a           &  \\
         & Halos  :  n/a        				&     Input:   $P(k)$ and $\bar{n}$    &  \\
\hline
\end{tabular}
\end{center}
\caption{Name of the methods, type of algorithm, halo definition, computing requirements and references for the compared methods. All computing times are given in cpu-hours per run and memory requirements are per run. We do not include the generation of initial conditions. The computational resources for halo finding in the N-body and \icecola mocks are included in the requirements. The computing time refers to runs down to redshift 1 except for the N-body where we report the time down to redshift 0 (we estimate an overhead of $\sim50\%$ between $z=1$ and $z=0$). Since every code was run in a different machine the computing times reported here are only indicative. We include the information needed for calibration/prediction of the covariance where relevant. ($^*$) In order to resolve the lower mass halos of the first sample a higher resolution version of \peakpatch should be run, requiring more computational resources than quoted here.}
\label{methods_tab}
\end{table*}

\subsection{Reference N-body: Minerva}

Our benchmark for the comparison is the set of N-body simulations called Minerva, first described in \citet{GriebEtal2016}. For this project $200$ additional realisations were run, for a total of $300$ realisations. These are performed using the code {\sc Gadget-3} \citep{Springel2005}, with $N_p=1000^3$ particles in a box of linear size $L_{box}=1500 \Mpc$ and a starting redshift of $z_{ini}=63$. The mass resolution is therefore $2.67\times10^{11}\Msun$. The initial conditions are given by the 2LPTic code\footnote{http://cosmo.nyu.edu/roman/2LPT/} and the cosmological parameters are fixed to their best fit value from the WMAP + BOSS DR9 analysis \citep{SanchezEtal2013}. The halo catalogues are obtained with the Subfind algorithm \citep{SpringelEtal2001}.

\subsection{ICE-COLA}
{\sc COLA} is a fast N-body method, that solves for particle trajectories using a combination of second order Lagrangian Perturbation Theory (2LPT) and a Particle-Mesh (PM) algorithm, where 2LPT is used for integrating the large scale dynamics while the PM part solves for the small scale dynamics \citep{TassevZaldarriagaEisenstein2013}. This allows to drastically reduce the number of time-steps needed by the solver to recover the large-scale clustering of a full N-body method within a few per-cent. In this article we use the \icecola version of the code, as optimised in \cite{IzardCrocceFosalba2016}. Thus, the mocks used in this work use $30$ time-steps linearly spaced in the expansion factor $a$ from the starting redshift at $z_{ini}=19$ to $z=0$\footnote{Only half of these steps are completed by $z=1$} and a PM grid-size of $3 \times N_p^{(1/3)}$ in each spatial dimension. They share the same particle load and box-size as Minerva. Note that this configuration is optimised for reproducing dark-matter clustering and weak lensing observables \citep{2015arXiv150904685I} and it is somewhat over-demanding in terms of computational cost for halo clustering alone \citep{IzardCrocceFosalba2016}, as seen in Table~\ref{methods_tab}. Halos are found on-the-fly using a Friends of Friends (FoF) algorithm with a linking length of $b=0.2$ the mean inter-particle distance. This method does not require calibration so we consider it a predictive method in the following.

\subsection{PINOCCHIO}
In this work we use the latest version of \pinocchio, presented in \citet{MunariEtal2017B}. The algorithm can be summarised as follows: first, a linear density field is generated, using $1000^3$ particles, and it is smoothed on different scales. For each particle the collapse time at all smoothing scales is computed using the ellipsoidal collapse in the 3LPT approximation and the earliest is assigned as collapse time of the particle. The process of halo formation is implemented through an algorithm that mimics the hierarchical assembly of dark matter halos through accretion of matter and merging with other halos. Finally, halos are displaced to their final position using 3LPT. This method only needs to be calibrated once and no re-calibration was required for this work, so we consider it predictive in what follows.

\subsection{PEAKPATCH}
In this work we use a new massively parallel version of the \peakpatch algorithm \citep{Steinetal2018,Alvarezinprep}, originally developed by \citet{BondMyers1996A, BondMyers1996B, BondMyers1996C}. This approach is a Lagrangian space halo finder that associates halos to regions that have just collapsed by a given time. The algorithm used in this work features four main steps: first, a density field is generated with $1000^3$ particles using the 2LPTic code. Collapsed regions are then identified using the homogeneous ellipsoidal collapse approximation and once exclusion and merging are imposed in Lagrangian space the regions are displaced to their final position using 2LPT. This method does not require calibration, so we will consider it a predictive method in the following.

\subsection{HALOGEN}
The \halogen method relies on the generation of a 2LPT matter density field at the redshift of interest and a subsequent assignment of halos according to an exponential bias prescription. The algorithm used in this work can be summarised as follows: first, a 2LPT matter density field is generated at $z=1$ using $768^3$ particles, downsampling the initial conditions of the Minerva simulations. The particles are then assigned to a $300^3$ grid. Halos are assigned to grid-cells with a probability proportional to $\rho_{cell}^{\alpha(M_h)}$, where $M_h$ is a halo mass sampled from the average mass function of the Minerva simulations. The position of the halo is drawn randomly from particle positions within the grid cell, while the velocity is given by the particle velocity multiplied by a velocity bias factor $f_{vel}(M_h)$. The function $\alpha(M_h)$ is calibrated using the average Minerva 2-point correlation function while the velocity bias one, $f_{vel}(M_h)$, is calibrated using the variance of halo velocities in Minerva, for several bins in halo mass. Therefore we consider \halogen as a calibrated method in the following.

\subsection{PATCHY}
In this work we employ the \patchy version described in \citet{KitauraEtal2015}, that uses augmented LPT (ALPT) to evolve the initial density field to the redshift of interest and then assigns halos using a non-linear scale-dependent and stochastic biasing prescription. The algorithm used in this work can be summarised as follows: first, a linear matter density field is generated using $500^3$ particles in a $500^3$ grid, downsampling the initial conditions of the Minerva simulations, and then evolved to $z=1$ using ALPT. The number density of halos is determined by using a deterministic bias prescription and halos are assigned to grid cells by sampling a negative binomial distribution, with a stochastic bias parametrisation to model deviations from Poissonianity. Halo velocities are given by modelling the coherent flow with ALPT and adding a Gaussian dispersion, while masses are drawn from the Minerva simulations and assigned according to halo environment properties (local density and cosmic web type) as determined by the HADRON code \citep{Zhao:2015jga}. Two separate set of mocks were produced for this work in order to reproduce the two samples considered (see Section \ref{sec:samples}). \patchy will be considered a calibrated method in the following.

\subsection{Lognormal}
The lognormal mocks used in this work are produced with the code described in \citet{AgrawalEtal2017}. The algorithm used in this work can be summarised as follows: a Gaussian field $G(\vec{x})$ is generated on a $256^3$ grid for matter and halos. For halos, we first measure the average real-space two-point correlation function $\xi$ in Minerva, which is ``gaussianised'' through the relation $\xi_G=\ln(1+\xi(r))$. We then get the power spectrum $P_G$ such that its Fourier transform matches $\xi_G$. This $P_G$ is used to generate the Gaussian density field in each point of the Fourier Space grid (where modes are independent). This field is then transformed back to configuration space as $G(\vec{x})$. In turn the matter density field is generated directly in Fourier space using the average power spectrum of Minerva and is only used to estimate the matter velocity field (needed to account for redshift space distortions) by solving the linear continuity equation. Both fields are then transformed into lognormal fields using the measured variance of the Gaussian fields, as $\delta_{lg}(\vec{x})=\exp{\left[-(1/2) \sigma^2_G+G(\vec{x})\right]}-1$, and added. The number of halos in a given cell is sampled from a Poisson distribution with mean $\bar{n}_h [ 1+\delta_h(\vec{x})] V_{cell}$, where $\bar{n}_h$ is the mean number density of halos in the Minerva simulations, $\delta_h(\vec{x})$ is the halo density field and $V_{cell}$ is the volume of the cell. Finally, halos are randomly positioned within the cell and their velocity is given by the velocity of the cell as computed by solving the linear continuity equation for matter. We will consider the lognormal mocks as an analytical prediction and use 1000 mocks to reduce the noise. Note that none of these are matched to the Minerva ICs.

\subsection{Gaussian covariance}\label{sec:gauss}
Here we use the Gaussian covariance model described in \citet{GriebEtal2016}. In this model the contribution of the trispectrum and of the super-sample covariance are neglected, so that the covariance is diagonal. The binned power spectrum multipole covariance in the thin shell approximation is given by:
\begin{multline}\label{eq:gauss_cov}
C_{\ell_1,\ell_2}(k_i,k_j)=\delta_{ij}\frac{(2\ell_1+1)(2\ell_2+1)}{2N_{k_i}} \\
\int^1_{-1} \left[ P(k_i,\mu) + \frac{1}{\bar{n}}\right]^2 \mathcal{L}_{\ell_1}(\mu) \mathcal{L}_{\ell_2}(\mu) \text{d}\mu,
\end{multline}
where $\delta_{ij}$ is the Kroneker delta function, $N_{k_i}$ is the number of independent Fourier modes in the bin, $P(k,\mu)$ is the anisotropic halo power spectrum, $\mu=\cos{\theta}$ where $\theta$ is the angle between the line-of-sight (LOS) and the separation vector of a galaxy pair and $\mathcal{L}_\ell$ denotes the Legendre polynomial of order $\ell$. In this work $P(k,\mu)$ is obtained by fitting the average halo power spectrum of the N-body simulations with the model given in Section~\ref{sec:model}. The evaluation of the covariance takes around 2 seconds on a regular computer.

%%%%%%%%%%%%%%%%%%%%%%%%%%%%%%%%%%%%%%%%%%%%%%%%
% Samples 	                               %
%%%%%%%%%%%%%%%%%%%%%%%%%%%%%%%%%%%%%%%%%%%%%%%%

\section{Halo Samples}
\label{sec:samples}
We define two samples in the N-body simulations by cutting in mass at two thresholds:  $1.1\times 10^{13}$ $M_{\odot}/h$ and $2.7\times 10^{13}$ $M_{\odot}/h$, that correspond to 42 and 100 particles respectively. This yields samples with number densities of $2.13\times10^{-4}\kvecMpccube$ and $5.44\times10^{-5}\kvecMpccube$ respectively. We only consider redshift $z=1$.

\begin{figure}
\begin{center}
\includegraphics{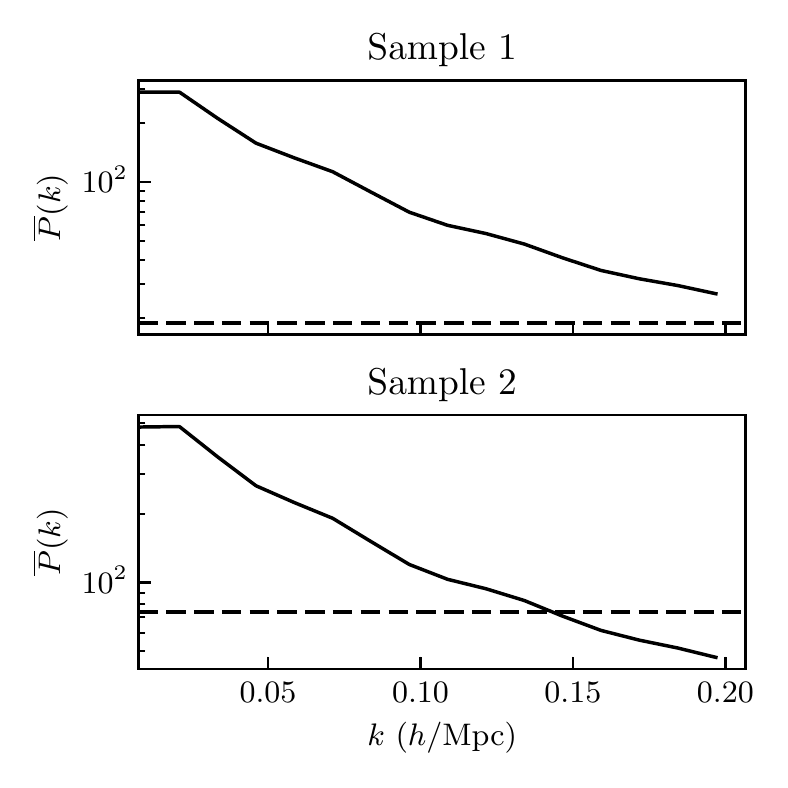}
\caption{Average real space halo power spectrum from the 300 N-body simulations (continuous line) and the corresponding Poisson shot noise level (dashed line), for sample 1 (top) and sample 2 (bottom) as defined in the top entries of Table~\ref{samples}, at redshift $1$.}
\label{minerva_rs}
\end{center}
\end{figure}

Results for the higher mass limit are more robust because the halos are sampled with a larger number of particles making the estimation of their mass and position more reliable. Nevertheless future surveys will observe galaxies that reside in lower mass halos and that have a higher number density, thus probing a different shot-noise regime. That is why we consider a sample at lower mass that is closer to the target population of such surveys (e.g. for \textit{Euclid} we expect a number density of $9\times10^{-4}\kvecMpccube$ at this redshift). In Figure~\ref{minerva_rs} we show the shot-noise subtracted average halo power spectrum in real space, with the corresponding level of shot noise indicated with a dashed line. While Sample 1 is always signal dominated in the range of scales that we consider, Sample 2 is shot-noise dominated for $k> 0.14 \kvecMpc$.

\begin{figure}
\begin{center}
\includegraphics{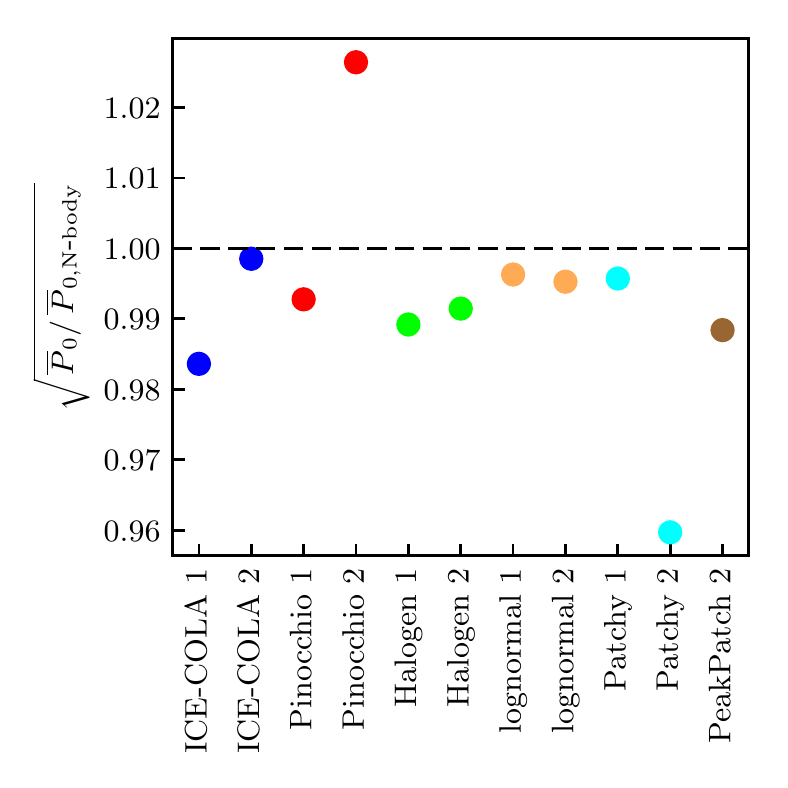}
\caption{Square root of the average clustering amplitude of the approximate mocks divided by the average of the N-body simulations in the range $0.008<k (\kvecMpc)<0.096$ for the two samples, as indicated in the labels.}
\label{bias_samples}
\end{center}
\end{figure}

We define equivalent samples in the approximate mocks by matching the abundance of halos to the average abundance in the N-body simulations. The calibrated methods have the same abundance of halos at these thresholds by construction, while in the case of the predictive methods we match the abundance by changing the mass threshold. The resulting samples are listed in Table~\ref{samples}. The abundance matching procedure does not yield exactly the same number of halos because the mass function is discretised in steps corresponding to the particle mass, nevertheless the recovered abundances agree within $1\%$. In the case of \pinocchio, since the abundance could not be matched to high precision, halo masses were made continuous by randomly distributing the masses of $N$-particle halos between $N \times M_p$ and $(N+1) \times M_p$, where $M_p$ is the particle mass of the Minerva simulations. The \peakpatch mocks, as implemented here, do not resolve the lower mass halos, hence results are presented only for the second sample. In Figure~\ref{bias_samples} we show the average clustering amplitude in the range $0.008<k \,(\kvecMpc)<0.096$ for the approximate mocks divided by the average of the N-body simulations in the same range, from which we can infer that the linear bias is recovered within a few percent in all the samples.

This procedure to define samples resembles what is done in the analysis of real data, where mocks are produced by matching the number density and bias of the observed sample. Moreover, since we expect shot-noise to have an important role in the covariance of these samples, matching the abundance will yield similar shot-noise levels in the compared covariances and will enhance discrepancies arising from other factors, notably from the approximations introduced by the different methods. Finally, since every approximate method defines halos differently it is not expected that cutting at the same mass threshold as the N-body simulations would yield the same halo samples, we therefore believe that samples matched by abundance will result in a fairer comparison. In Appendix~\ref{appendix1} we compare results for the predictive methods for sample defined by cutting at the same mass threshold and samples matched by abundance. The abundance matching procedure is especially helpful for the \peakpatch mocks in which halos are identified as lagrangian spherical overdensities, contrary to the other methods that use friends-of-friends (FoF) or are calibrated to reproduce FoF mass functions.

\begin{table}
\begin{center}
\begin{tabular}{l|c|c}
Method & ${\bar n}_{halos}$ ($h^{3}$Mpc$^{-3}$) & $M_{min}$ ($\Msun$) \\
%Sample 1
\hline
\multicolumn{3}{c}{Sample 1} \\
\hline 
N-body & $2.130\times10^{-4}$& $1.121 \times 10^{13}$ \\
\icecola & $2.123\times10^{-4}$ & $1.086 \times 10^{13}$ \\
\pinocchio & $2.148\times10^{-4}$ & $1.044 \times 10^{13}$ \\
\halogen & $2.138\times10^{-4}$& $1.121 \times 10^{13}$ \\
Lognormal & $2.131\times10^{-4}$ & $1.121 \times 10^{13}$ \\
\patchy & $2.129\times10^{-4}$ & $1.121 \times 10^{13}$ \\
%Sample 2
\hline
\multicolumn{3}{c}{Sample 2} \\
\hline
N-body & $5.441\times10^{-5}$ & $2.670 \times 10^{13}$ \\
\icecola & $5.455\times10^{-5}$ & $2.767 \times 10^{13}$ \\
\pinocchio & $5.478\times10^{-5}$ & $2.631 \times 10^{13}$ \\
\halogen & $5.393\times10^{-5}$ & $2.670 \times 10^{13}$ \\
Lognormal & $5.441\times10^{-5}$ & $2.670 \times 10^{13}$ \\
\patchy & $5.440\times10^{-5}$ & $2.670 \times 10^{13}$ \\
\peakpatch & $5.439\times10^{-5}$ & $2.355 \times 10^{13}$ \\
\end{tabular}
\end{center}
\caption{Characteristics of the halo samples considered. For Sample 1
  (top) we define in each approximate method an equivalent sample
  matched by abundance to the density of halos more massive than
  $1.12\times 10^{13}\Msun$ in
  the N-body simulation. For Sample 2 we do equivalently, but matching abundance for
  halos with $M_h \ge 2.67\times 10^{13} \Msun$ in the N-body. Second column displays the resulting
  abundance of those samples, which are not exactly equal to the
  N-body due to mass function discretisation in steps equivalent to the particle mass. Samples are defined at $z=1$. The significant difference in the mass cut of \peakpatch is due to the different mass definition criterium, as explained in Appendix \ref{appendix1}.}
\label{samples}
\end{table}

%%%%%%%%%%%%%%%%%%%%%%%%%%%%%%%%%%%%%%%%%%%%%%%%
% Methodology				       %
%%%%%%%%%%%%%%%%%%%%%%%%%%%%%%%%%%%%%%%%%%%%%%%%

\section{Methodology}
\label{sec:methodology}

We compare 300 realisations for each approximate method with initial conditions matching the Minerva N-body simulations, so that the comparison is not affected by sample variance.

\subsection{Clustering measurements}
We compute the redshift space power spectrum multipoles in the distant-observer approximation. We choose the direction of the LOS parallel to the three axes of the simulation box and average the results for the three directions to further reduce the noise in the measurements. We compare results at redshift $z = 1$ and in the range of scales $0.008<k\, (\kvecMpc)<0.197$, unless otherwise stated. We measure the power spectrum multipoles $P_\ell$ ($\ell=0,2,4$) using the public code PowerI4\footnote{https://github.com/sefusatti/PowerI4}, that employs the interlacing technique to reduce aliasing contribution \citep{SefusattiEtal2016}. We use the 4th order mass interpolation scheme (Piecewise Cubic Spline, PCS) on a $256^3$ grid. We bin the power spectrum multipoles in intervals of $\Delta k=3 k_f$, where $k_f=2\pi/L_{box}$ is the fundamental frequency of the box, for a total of 16 bins per multipole.

\subsection{Theory modelling and parameter space}\label{sec:model}

We compare the performance of different covariance matrices in terms of the errors and contours in a series of model parameters which are generally standard in galaxy clustering analysis. In particular we adopt a model very close to the one in the anisotropic clustering study of BOSS DR12 data presented in \cite{2017MNRAS.467.2085G} 
and \cite{SanchezEtal2017b}. Such model includes: (1) nonlinear matter clustering through the matter and velocity divergence correlations (gRPT, Crocce et al. in prep.), (2) a biasing scheme to one-loop in the initial power spectra (3) the non-linear equivalent of the Kaiser effect for redshift space distortions through the correlations between densities and large-scale velocities. The main difference with respect to a galaxy analysis is that we do not consider parameters related to small scale pair-wise velocity dispersion (the so called Fingers-of-God effect), because we are considering halos and these are not affected by this effect. Broadly speaking, the theory model is given by:
\begin{eqnarray}
\label{eq:rsd_th}
P_{hh}^s(k,\mu) & = & P_{hh}(k) + 2 f \mu^2 P_{h\theta}(k) + f^2 \mu^4 P_{\theta\theta}(k) \nonumber \\ 
               & + &  P^{(2)}_\text{novir}(k,\mu) + P^{(3)}_\text{novir}(k,\mu)
\end{eqnarray}
where $P_{hh}$, $P_{h\theta}$ and $P_{\theta\theta}$ stand for halo density $h$ and velocity divergence $\theta$ auto and cross power spectra, and $f$ is the growth rate of structure (where $\theta \equiv \nabla {\bf u} \equiv
 - \nabla {\bf v} / a H f$). The first three terms are the non-linear ``Kaiser'' effect, with $P_{hh}$ and $P_{h\theta}$ written to one-loop in the biasing scheme. Explicit expressions can be found in \cite{SanchezEtal2017b} and they depend on $b_1$ (linear bias), $b_2$ (non-linear, second order in density fluctuations), $\gamma_2$ (non-local, second order) and $\gamma^{-}_3$ (non-local, third order in perturbations), following the notation and bias basis of \cite{2012PhRvD..85h3509C}. Higher order terms, such as $b_3$, are ``renormalised'' into the above parameters (at one-loop), see \cite{2006PhRvD..74j3512M}.
 The last two terms in Eq.~(\ref{eq:rsd_th}) also arise from the transformation of real to redshift space coordinates.
Defining $D_s \equiv \delta_h + f \nabla_z u_z$, we have
\begin{eqnarray}
P_{\rm novir}^{(2)}(k,\mu) = \int {q_z\over q^2} \Big[ B_{\theta D_s D_s}(\vq,\vk-\vq,-\vk) \nonumber  \\ \quad + B_{\theta D_s D_s}(\vq,-\vk,\vk-\vq)\Big] d^3q, \label{Pnovir2}
\end{eqnarray}
which depends on the cross bispectrum $B$ between the velocity divergence $\theta$ and $D_s$,
which can be evaluated at tree-order (thus, depends only on $b_1$, $b_2$ and $\gamma_2$), while
\begin{eqnarray}
P_{\rm novir}^{(3)}(k,\mu) = \int {q_z\over q^2} {(k_z-q_z)\over (\vk-\vq)^2} (b_1+f \mu_q^2)(b_1+f \mu_{k-q}^2) \nonumber \\
                           \quad \times P_{\delta\theta}(k - q)P_{\delta\theta}(q) d^3q,  \label{Pnovir3}
\end{eqnarray}
which is already second-order in $P(k)$ and hence only depends on linear bias $b_1$. Equations (\ref{Pnovir2}) and (\ref{Pnovir3}) are basically the equivalent of the $A(k,\mu)$ and $B(k,\mu)$ terms in Eq. (18) of \cite{TNS} extended to biased tracers (see also \cite{eTNS}). 
We have checked that the sensitivity of our halo samples clustering to $\gamma_2$ is minor, hence we keep terms proportional to $\gamma_2$ but we fix $\gamma_2$ to its local lagrangian relation to the linear bias, i.e. $\gamma_2 = -(2/7)(b_1-1)$.  

The final component of the model is the Alcock-Paczynski effect. Anisotropic clustering analysis needs to assume a ``fiducial'' cosmological model in order to transform observables (redshifts and angular positions) into co-moving coordinates. A mismatch between this ``fiducial'' model and the ``true'' cosmology (i.e. the one being assumed in the likelihood evaluation) leads to distortions, known as Alcock-Paczynski (AP) effect \citep{AP}, that can be used to place cosmological constrains \citep[e.g.][]{2012MNRAS.420.2102S}. Such distortions are characterised by two parameters that transforms comoving coordinates in the ``true'' cosmology, denoted $(k_\perp,k_\parallel)$\footnote{Here $\perp$ and $\parallel$ refer to wave-vectors parallel and perpendicular of the LOS.}, to 
coordinates in the ``fiducial'' cosmology, denoted $(k_\perp^\prime,k_\parallel^\prime)$, as
\begin{equation}
k^\prime_\perp = \alpha_\perp k_\perp \quad k^\prime_\parallel = \alpha_\parallel k_\parallel ,
\end{equation}
with
\begin{equation}
\alpha_\perp = \frac{D_A(z)r^\prime_d}{D_A^\prime(z) r_d} \quad \alpha_\parallel = \frac{H^\prime(z) r^\prime_d}{H(z) r_d} ,
\end{equation}
where $D_A$ is the angular diameter distance to the sample redshift, $H(z)$ the Hubble parameter and $r_d$ is the sound horizon at the drag redshift. In terms of wave-vector
amplitude and angle to the line-of-sight this transformation is \citep{1996MNRAS.282..877B},
\begin{eqnarray}
k(\mu^\prime,k^\prime) & = & k^\prime \left[ \alpha^{-2}_\parallel {\mu^{\prime}}^2 + \alpha^{-2}_\perp (1-{\mu^\prime}^2) \right]^{1/2} \\
\mu(\mu^\prime) & = & \mu^\prime \alpha^{-1}_\parallel \left[ \alpha^{-2}_\parallel {\mu^{\prime}}^2 + \alpha^{-2}_\perp (1-{\mu^\prime}^2) \right]^{-1/2}.
\end{eqnarray}
Therefore, after evaluating the power spectrum model at the cosmology being assumed by the likelihood step, i.e. $P(k,\mu)$, we need to transform back to the basis of the ``fiducial'' cosmology, together with the multipole decomposition, as 
\begin{equation}
P_\ell (k^\prime)  = \frac{2\ell+1}{2 \alpha^2_\perp \alpha_\parallel} \int   \mathcal{L}_\ell(\mu^\prime) P(k(k^\prime,\mu^\prime),\mu(\mu^\prime)) d\mu^\prime.
\end{equation}
In summary, our parameter space has six free parameters: $b_1$, $b_2$, $\gamma^{-}_3$ for galaxy physics, the growth rate of structure $f \sigma_8$, and the two AP parameters $\alpha_\perp$, $\alpha_\parallel$. 

\subsection{Covariance Matrix Estimation}

The covariances are computed using the sample covariance estimator:
\begin{equation}
C_\ell(k_i,k_j)=\frac{1}{N_s-1}\sum_{n=1}^{N_s} (P^n_\ell(k_i)-\bar{P}_\ell(k_i))(P^n_\ell(k_j)-\bar{P}_\ell(k_j)),
\end{equation}
where $N_s=300$ is the number of simulations, $\ell=0,2,4$ is the multipole considered and $\bar{P}_\ell=1/N_s\sum_{n=1}^{N_s}P_\ell^n$ is the average power spectrum. Analogously, the cross-covariances between different multipoles are computed as:
\begin{multline}
C_{\ell_1,\ell_2}(k_i,k_j)=\frac{1}{N_s-1}\sum_{n=1}^{N_s} (P^n_{\ell_1}(k_i)-\bar{P}_{\ell_1}(k_i)) \\ (P^n_{\ell_2}(k_j)-\bar{P}_{\ell_2}(k_j)).
\end{multline}
In what follows we compute covariance matrices in each of the three LOS directions independently and then average them.

\subsection{Fitting procedure}\label{sec:fitting}

The main use for covariances in the context of galaxy survey data analysis is to perform a fit to the measured observable given a model in the likelihood analysis framework. In order to understand if the approximations in the methods presented here systematically increase/decrease the errors on the recovered parameters we run MCMC chains based on the assumption of a Gaussian likelihood:
\begin{multline}
-2\log\mathcal{L}=\sum_{\ell_1\ell_2}\sum_{ij}(\hat{P}_{\ell_1}(k_i)-P_{m,\ell_1}(k_i))^T \\ \Psi_{\ell_1\ell_2}(k_i,k_j) (\hat{P}_{\ell_2}(k_j)-P_{m,\ell_2}(k_j)),
\end{multline}
where $\Psi=C^{-1}$ is the precision matrix and $P_m$ is a model of the power spectrum multipoles that depends on the cosmological and nuisance parameters. 

Since we are only interested in the performance of the covariance we use as $\hat{P}$ a fit to the N-body average power spectrum multipoles with the model described in Section~\ref{sec:model}. To do this fit we fix the cosmological parameters to their true values and only vary the bias parameters $b_1$, $b_2$ and $\gamma^{-}_3$. The best fit parameter values that we obtain for the two samples are given in Table~\ref{tab:best_fit}.

\begin{table}
\begin{center}
\begin{tabular}{l|c|c|c}
Sample & $b_1$ & $b_2$ & $\gamma^{-}_3$ \\
\hline
1 & 2.637 & -1.660 & 0.693 \\
2 & 3.434 & -1.067 & 1.601 \\
\end{tabular}
\end{center}
\caption{Best fit parameter values to the N-body power spectrum multipoles for the two samples.}
\label{tab:best_fit}
\end{table}

We then run MCMC chains by changing the covariance with the ones estimated with the approximate mocks. We vary the Alcock-Paczynski parameters ($\alpha_{\parallel}$,$\alpha_{\perp}$) and $f\sigma_8$ together with the nuisance bias parameters $b_1$, $b_2$ and $\gamma^{-}_3$. This matches what is done in article I for the configuration space analysis. For each method we first run a short chain to have an estimate of the parameter covariance and then run a full chain that uses that parameter covariance as input. All the full chains have $2\times10^5$ points.

%%%%%%%%%%%%%%%%%%%%%%%%%%%%%%%%%%%%%%%%%%%%%%%%
% Results				       %
%%%%%%%%%%%%%%%%%%%%%%%%%%%%%%%%%%%%%%%%%%%%%%%%

\section{Results}
\label{sec:results}

\subsection{Mean of the power spectrum multipoles}
\begin{figure*}
\begin{center}
\includegraphics[trim = 0cm 0 0cm 0, width=0.4\textwidth]{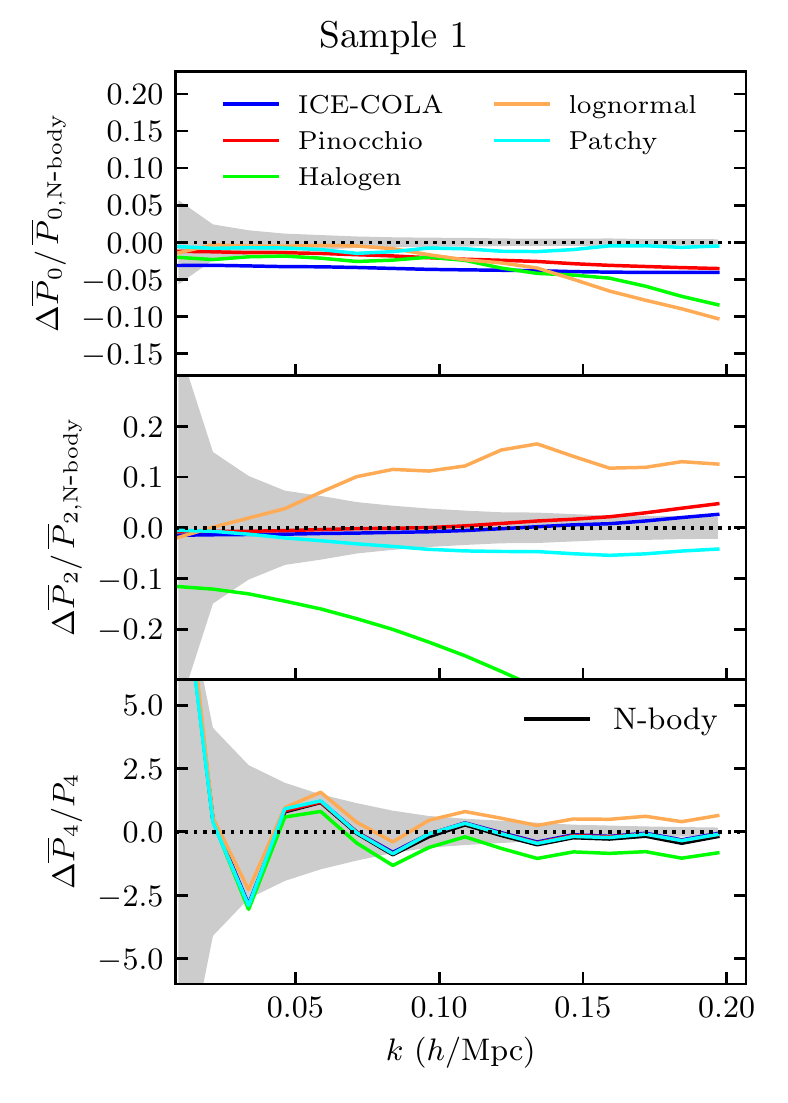}
\includegraphics[trim = 0cm 0 0cm 0, width=0.4\textwidth]{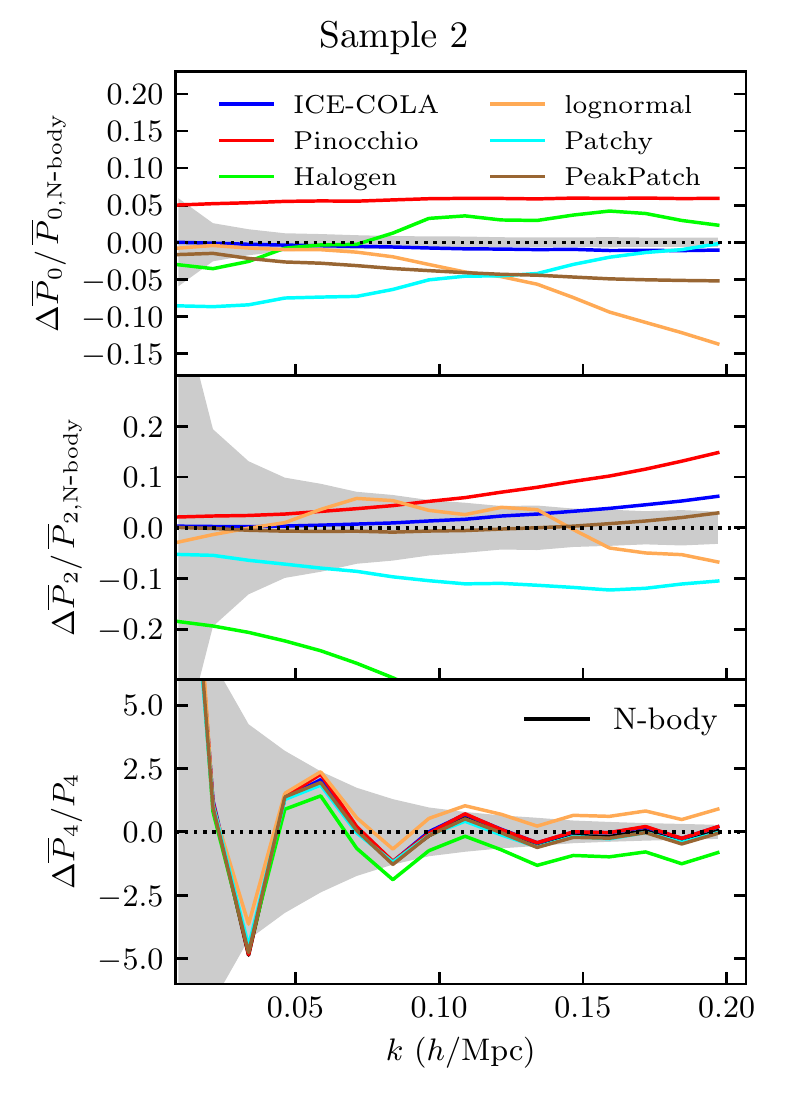}
\caption{Relative difference of the power spectrum monopole and quadrupole (hexadecapole) with respect to the N-body (model) ones for sample 1 (left plot) and sample 2 (right plot). In each plot we show the monopole (top panel), quadrupole (middle panel) and hexadecapole (bottom panel). The grey shaded region shows the standard deviation of the N-body measurements.}
\label{pkmlim}
\end{center}
\end{figure*}

Figure~\ref{pkmlim} shows the relative difference of the average power spectrum multipoles as measured from the approximate mocks with respect to the corresponding N-body measurements, except for the case of the hexadecapole where we use the theoretical estimate as reference because the large noise makes the N-body measurements cross 0 and can lead to diverging ratios. The shaded regions show the standard deviation of the 300 N-body simulations. In general the monopole is better reproduced than the higher order multipoles by all methods and the agreement with the N-body results is better in the first sample than in the second.

The monopole is reproduced to within $\sim5\%$ by all methods at large scales in both samples, with the exception of \patchy in the second sample. \icecola and \pinocchio perform the best in terms of scale dependence showing a very flat difference, as well as \patchy in the first sample. In the second sample the calibrated methods show a scale dependence and oscillations in the ratio, indicating that the BAO features are not completely reproduced by these methods in this high-bias sample. \patchy in the first sample and \icecola in the second have the best performances: they reproduce the monopole to within $1\%$ on the entire range of scales considered here. Notice that flat residuals in the monopole indicate that there is a mismatch in the linear bias, which is considered a nuisance parameter in cosmological analysis.

All methods show some degree of scale dependent difference in the quadrupole, with \halogen performing notably bad even at large scales. The other methods show a good agreement at large scales that deteriorates at small scales. \icecola and \pinocchio have very good performances in the first sample, reproducing the quadrupole to within $1\%$ in the whole range of scales considered. The same is true for \peakpatch in the second sample, where \icecola has a residual deviation of $\sim 5\%$ by $k=0.2\Mpc$ and \pinocchio $\sim 14\%$. \patchy shows $5\%-10\%$ deviations in the quadrupole. This performance can be potentially improved with an explicit fit to the velocity dispersion, modelling virialised motions, in the N-body. Lastly, we note that lognormal shows residual oscillations at BAO scales also in the quadrupole for both samples. 

In the case of the hexadecapole it is more difficult to evaluate per-cent differences because of noise amplification due to the very small values of the denominator when taking the ratio, but these are on average of the order of $10-15\%$ for \icecola, \pinocchio, \peakpatch and \patchy while they are $>40\%$ for \halogen and Lognormal.

\subsection{Variance of the power spectrum multipoles}
\begin{figure*}
\begin{center}
\includegraphics[trim = 0cm 0 0cm 0, width=0.4\textwidth]{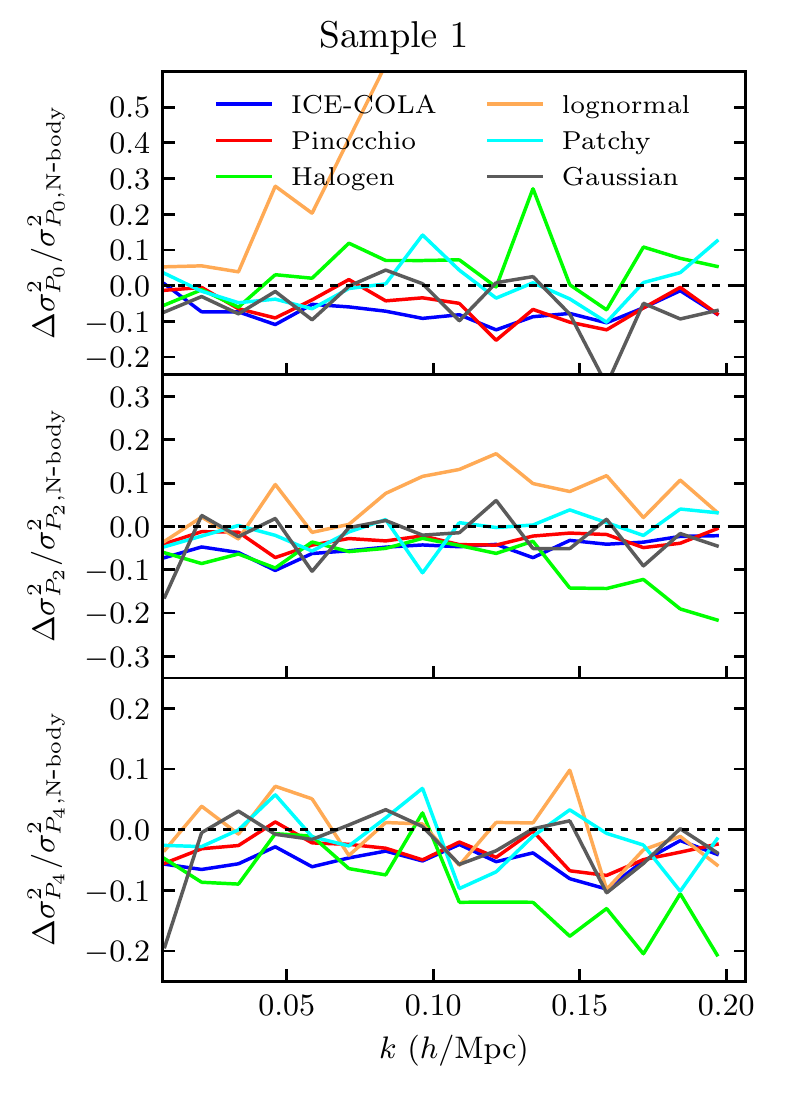}
\includegraphics[trim = 0cm 0 0cm 0, width=0.4\textwidth]{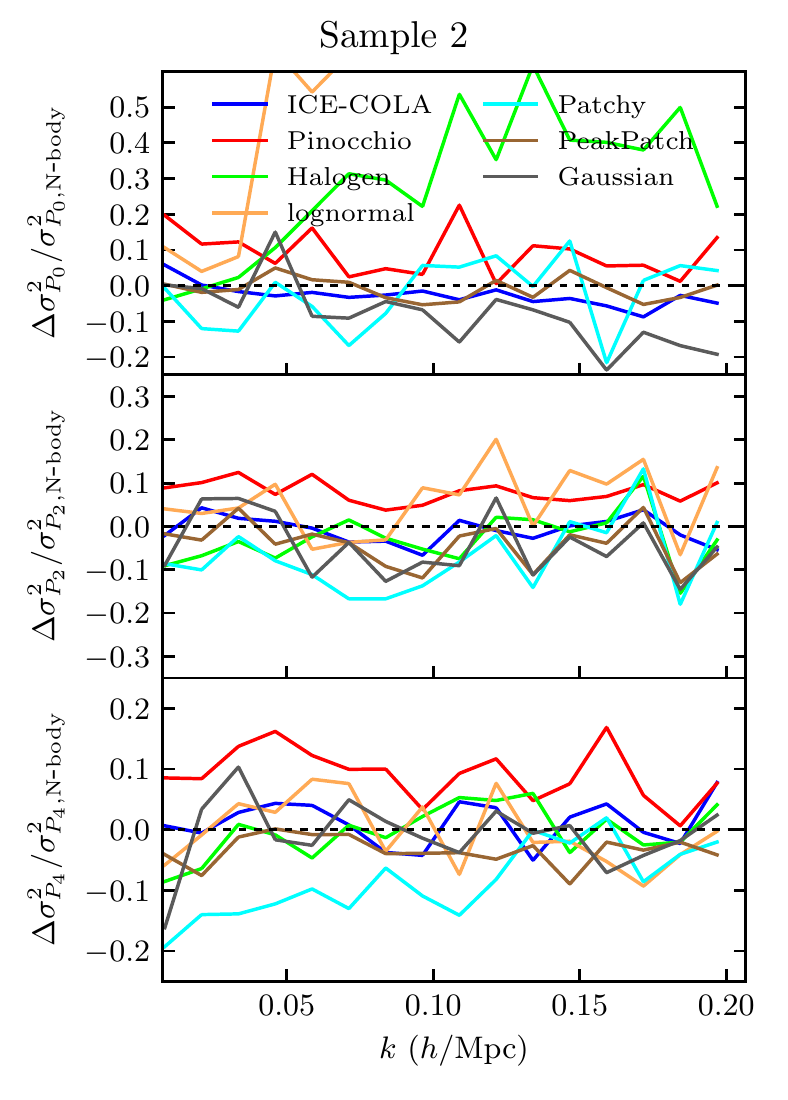}
\caption{Relative difference of the variance of the power spectrum multipoles with respect to the N-body ones for sample 1 (top plot) and sample 2 (bottom plot). In each plot we show the monopole (top panel), quadrupole (middle panel) and hexadecapole (bottom panel). Notice that the Gaussian is a noiseless estimate so that differences are dominated by the noise in the N-body variance estimate.}
\label{varmlim}
\end{center}
\end{figure*}

In Figure~\ref{varmlim} we show the relative difference of the variance of the power spectrum multipoles with respect to the N-body variance. We also include the Gaussian prediction described in Section~\ref{sec:gauss}. Note that the variance of the power spectrum in the Gaussian model is by construction noiseless hence the scatter in the plotted points is due to noise in the reference N-body variance (the denominator of the ratio). We notice that the variance of Sample 2 has a larger scatter, pointing to a larger contribution of shot-noise in the covariance error for this sample.

In Sample 1 all the methods agree with the N-body variance within $10\%$ on large scales, with the exception of Lognormal for reasons that we address below. \halogen reaches a $20\%$ difference beyond $k \sim 0.1\Mpc$. In Sample 2 \icecola and \peakpatch do particularly well with an average of $\sim3\%$ deviations from the N-body results in all multipoles, while the other methods show a more diverse behaviour. In the monopole variance \pinocchio has an average deviation of $\sim 9\%$ and \halogen of $30 \%$. The quadrupole variance is reproduced at the level of $5-10\%$ while the hexadecapole variance is within $5\%$ for all methods, with the exception of \pinocchio and \patchy ($9\%$). Overall the Gaussian model performs well, at the level of the mock based variances.

One noteworthy result is that Lognormal mocks show a significantly larger variance than the N-body in the monopole and that the discrepancy has a strong dependence on binning, showing increasing deviations with larger bins. We attribute this behaviour to the fact that these mocks show larger non-Gaussian contributions than the N-body simulations. In fact, the Gaussian part of the variance decreases with increasing bin widths while the non-Gaussian part does not depend on the binning \citep[see e.g.][]{ScoccimarroZaldarriagaHui1999}. Since here we are showing the relative difference, the relative magnitude of the non-Gaussian part increases with larger binning, inducing larger deviations from the N-body result. In addition, we found a trend of increasing deviations from the N-body with increasing bias that will be investigated in future work. All the other methods do not show this pronounced behaviour with binning, indicating that the relative contributions of the Gaussian and non-Gaussian parts to the covariance are better reproduced.

\subsection{Correlation coefficient of the power spectrum multipoles}
\begin{figure*}
\begin{center}
\includegraphics[trim = 2cm 0 2cm 0, width=\textwidth]{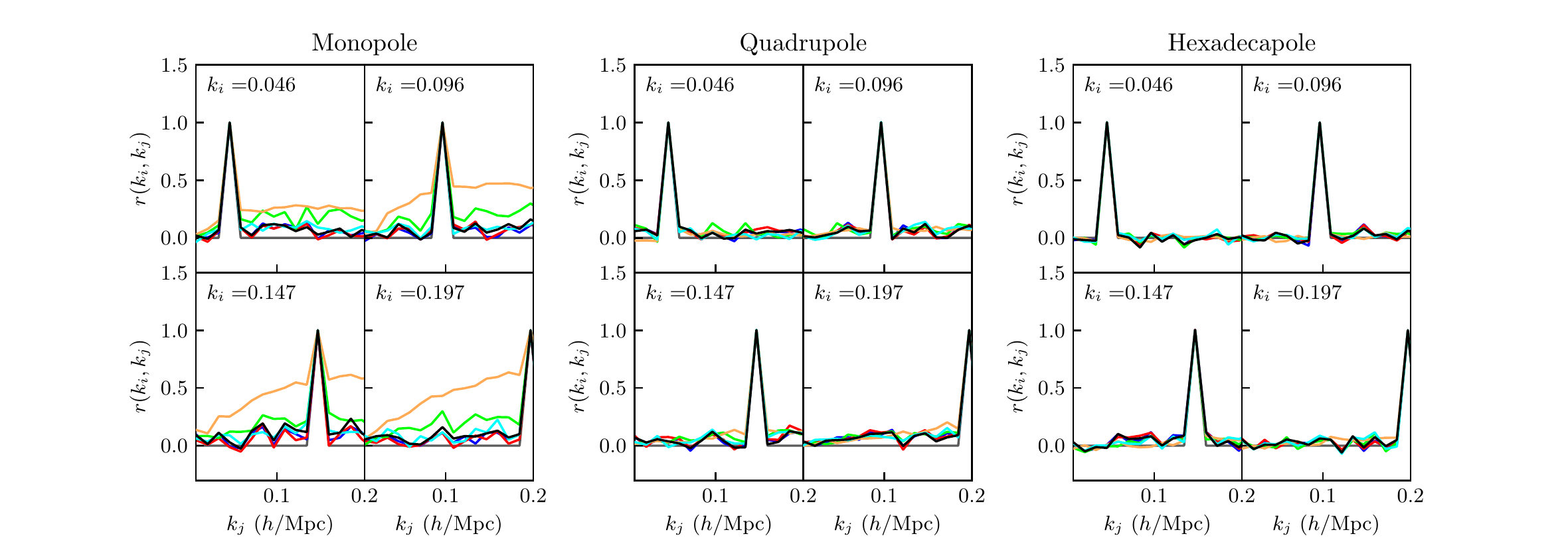}
\caption{Cut through the correlation coefficient of the monopole (left plot), quadrupole (middle plot) and hexadecapole (right plot) for sample 1 at four different values of $k$ as indicated in the panels. We show the results for N-body (black), \icecola (blue), \pinocchio (red), \halogen (green), \patchy (cyan), Lognormal (orange) and Gaussian (grey).}
\label{corr_coeff1}
\end{center}
\end{figure*}

\begin{figure*}
\begin{center}
\includegraphics[trim = 2cm 0 2cm 0, width=\textwidth]{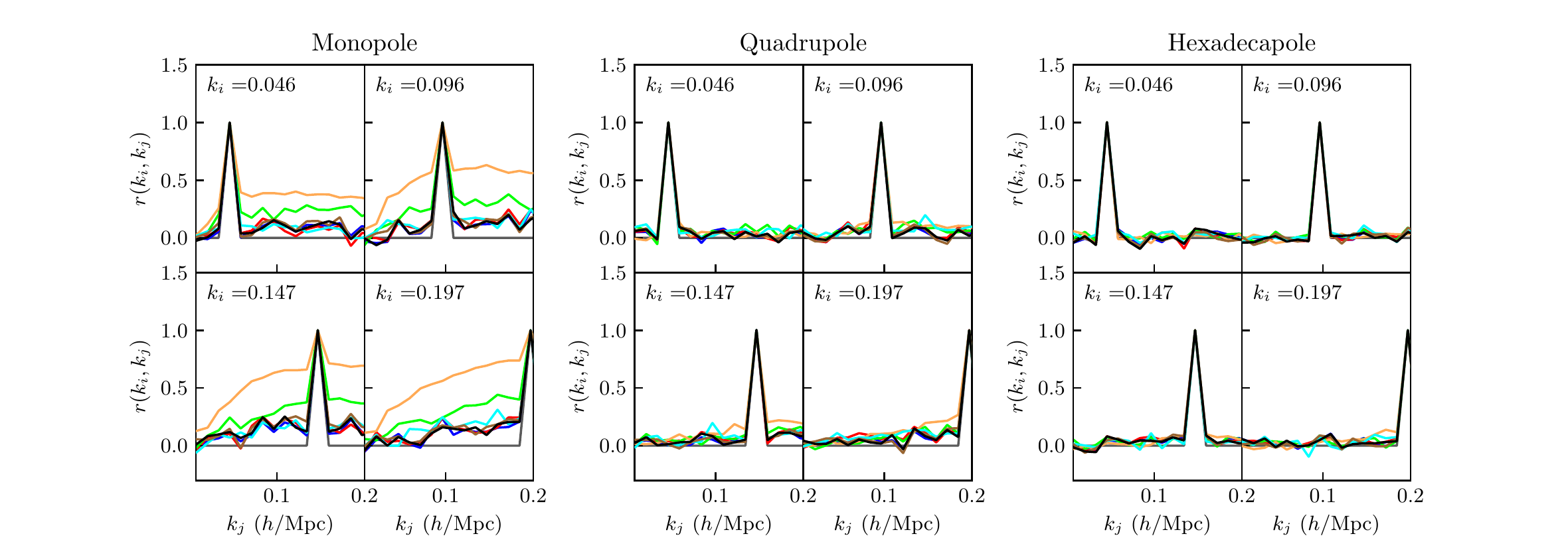}
\caption{Cut through the correlation coefficient of the monopole (left plot), quadrupole (middle plot) and hexadecapole (right plot) for sample 2 at four different values of $k$ as indicated in the panels. We show the results for N-body (black), \icecola (blue), \pinocchio (red), \peakpatch (brown), \halogen (green), \patchy (cyan), Lognormal (orange) and Gaussian (grey).}
\label{corr_coeff2}
\end{center}
\end{figure*}

To show the off-diagonal elements of the covariance matrix we compute the correlation coefficient for each method as
\begin{equation}\label{corr_coeff_M}
r_{ij}=\frac{C_\ell(k_i,k_j)}{\sqrt{C_\ell (k_i,k_i) C_\ell (k_j,k_j)}}.
\end{equation}
In Figures~\ref{corr_coeff1}-\ref{corr_coeff2} we show a cut through the correlation coefficient $r$ for the power spectrum multipoles for 4 different $k_i$ values, as indicated in the plot panels. The colour coding is the same as the previous plots with the addition of the N-body in black. The first consideration to make is that the level of correlation for the power spectrum multipoles at these scales and with the chosen binning is very low. The monopole covariance shows the largest correlations, on average $\sim7\%$ with a maximum of $\sim20\%$, while for the quadrupole they are on average of the order of $\sim4\%$ and for the hexadecapole $\sim1\%$. The second consideration is that thanks to the matching ICs all the approximate method show noise properties that are highly correlated to the noise in the N-body covariance. Finally, all methods seem to qualitatively reproduce the level of correlations of the N-body spectra with the exception of the monopole for \halogen, that predicts correlations $2$ to $3$ times higher than the N-body, and Lognormal for the reason explained in the previous section.

\subsection{Cosmological parameter constraints}
\begin{figure*}
\includegraphics[width=0.82\textwidth]{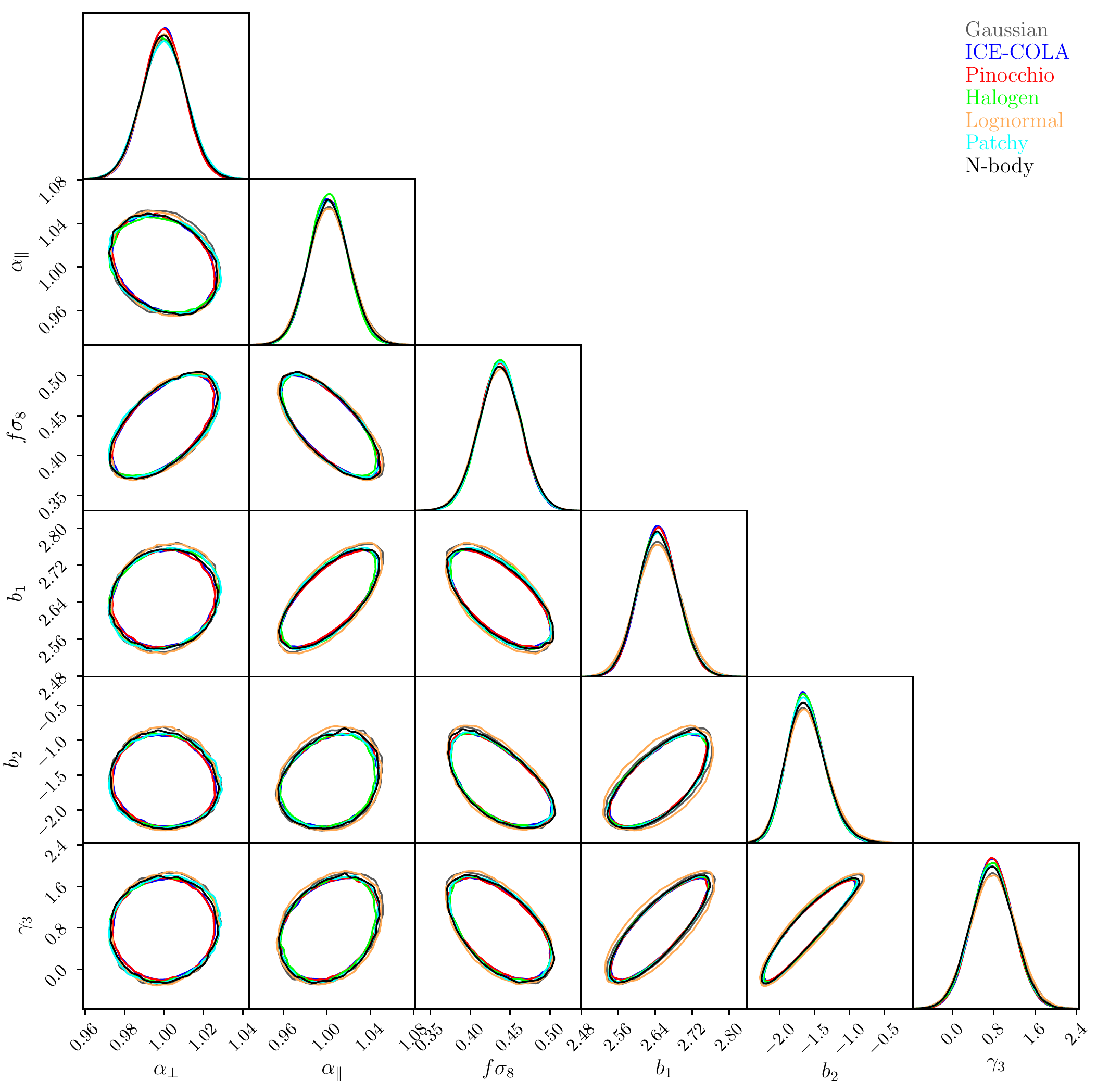}
\caption{Marginalised $2\sigma$ contours for cosmological and nuisance parameters for sample 1.}
\label{contour1}
\end{figure*}

\begin{figure*}
\includegraphics[width=0.82\textwidth]{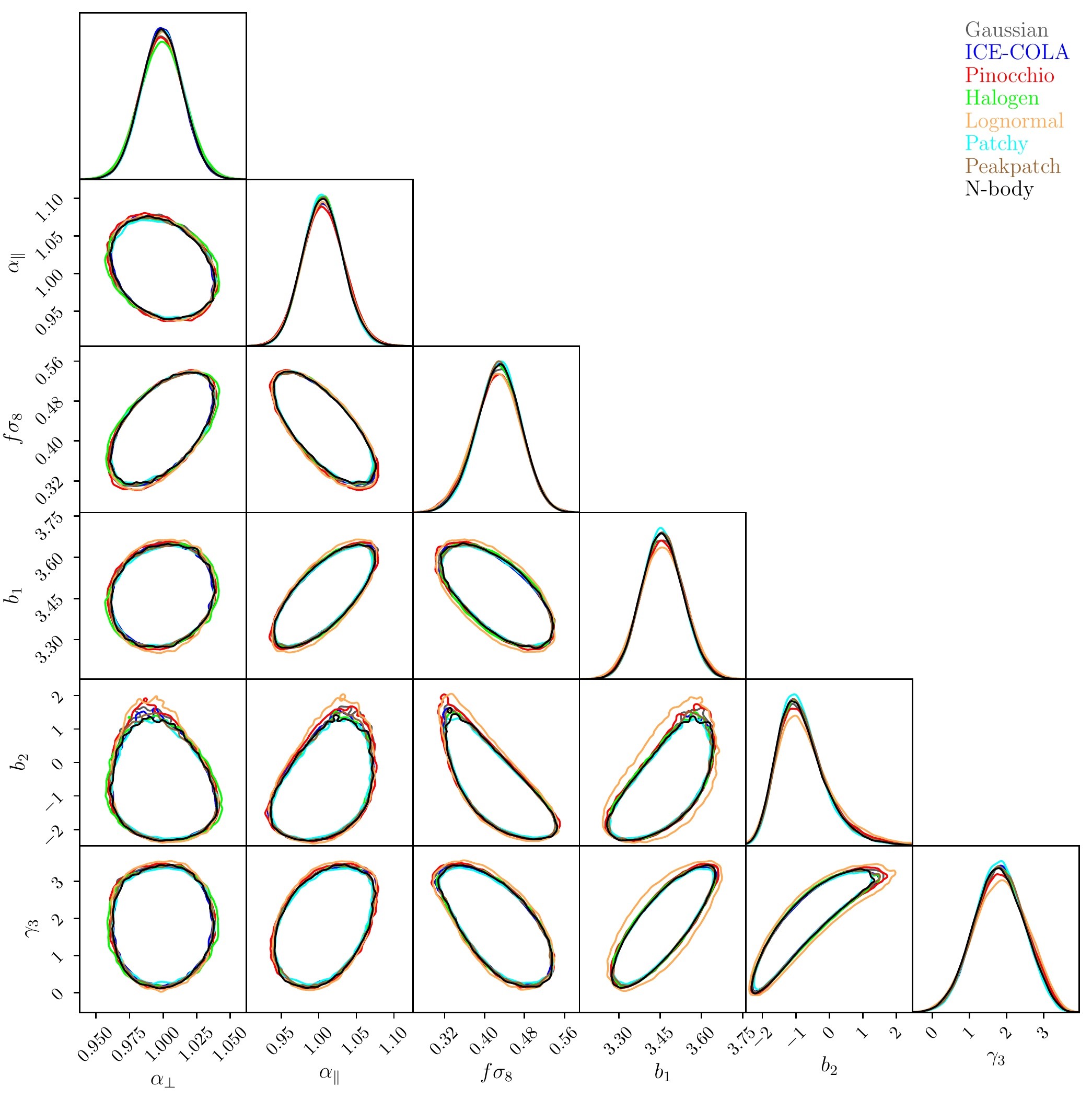}
\caption{Marginalised $2\sigma$ contours for cosmological and nuisance parameters for sample 2.}
\label{contour2}
\end{figure*}

To gauge the accuracy of the approximate methods in a realistic context we propagate errors on the covariance all the way to cosmological parameter errors using a likelihood analysis, as described in Section \ref{sec:fitting}. In Figures~\ref{contour1}-\ref{contour2} we show the resulting $2\sigma$ parameter contours for the cosmological and nuisance parameters. Overall the position and direction of degeneracies of the contours is well reproduced by the approximate mocks. Recovered best fit parameter values\footnote{Defined as the maximum of the posterior distribution for each parameter after all others have been marginalised over.}, for both samples, agree with the input ones to $\lesssim 1\%$ for all cosmological parameters and $b_1$, but show larger deviations for the higher order halo bias ones even when using the N-body covariance. We attribute this to the fact that the latter parameters are poorly constrained and sometimes show asymmetric posterior distributions (see Figs.~\ref{contour1},\ref{contour2}). Similar results are obtained if the best-fit values are defined as the mean of the marginalised posterior distribution. In any case, these best-fit parameter biases are always marginally negligible if defined with respect to their corresponding 1-$\sigma$ error.

\begin{figure}
\includegraphics{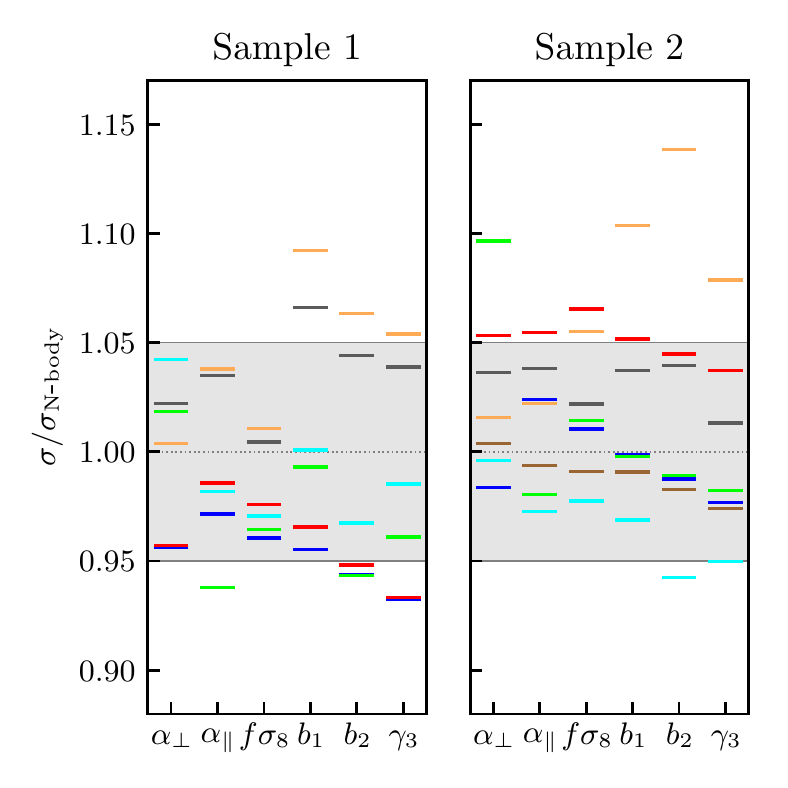}
\caption{Ratio of the error on cosmological and nuisance parameters with respect to N-body for sample 1 (left) and sample 2 (right). The grey shaded area indicate $5\%$ differences. We show the results for \icecola (blue), \pinocchio (red), \peakpatch (brown), \halogen (green), \patchy (cyan), Lognormal (orange) and Gaussian (grey).}
\label{err_ratio}
\end{figure}

To quantify differences between the recovered parameter errors we plot the ratio of the marginalised error on cosmological and nuisance parameters with respect to the N-body results in Figure~\ref{err_ratio}.  Before further discussion we point out that despite matching the ICs there is a residual scatter on parameter error ratios that we quantify in Appendix~\ref{appendix_vc} to be at the level of $4\%-5\%$, (indicated in Figure~\ref{err_ratio} as a grey shaded region). One should bear this in mind when evaluating differences found. Moreover, for lognormal and Gaussian the error exceeds this value because they do not match the ICs to the N-body simulations and thus have a different realisation of noise in the covariance. This gets propagated to the cosmological parameter errors. This means that we expect results for these two methods to be more strongly affected by the noise in the N-body covariance. We estimate uncertainties for these two methods to be at the level of $10\%$.

Overall all the methods reproduce the N-body errors within the statistical uncertainty, with few exceptions. \pinocchio slightly exceeds the $5\%$ limit in the nuisance parameters $b_2$ and $\gamma_3^-$ in Sample 1 and in the cosmological parameters and $b_1$ in Sample 2, while \icecola has similar issues for the nuisance parameters $b_2$ and $\gamma_3^-$ only in Sample 1. \halogen shows deviations larger than the uncertainty in one of the AP parameters in both samples, while \patchy only shows a slight underestimation of the error on $b_2$ in Sample 2. In Sample 2 Lognormal shows deviations for the nuisance parameters that are larger than the expected $10\%$ statistical uncertainty.

\begin{figure}
\includegraphics{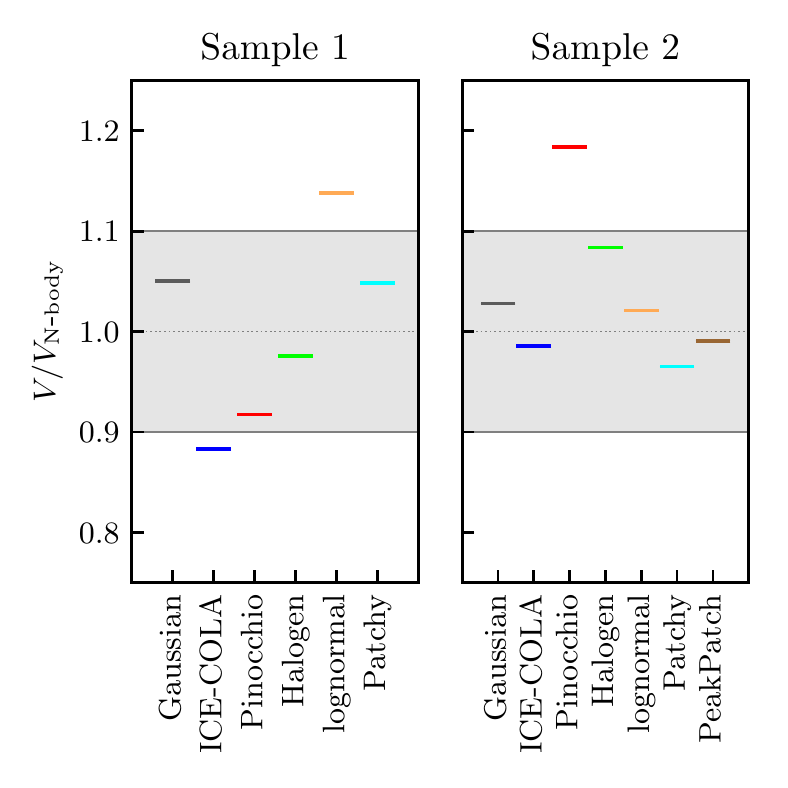}
\caption{Ratio of the volume of the 3D ellipsoid of cosmological parameters with respect to N-body for sample 1 (left) and sample 2 (right). The grey shaded area indicate $10\%$ differences.}
\label{vol_ell}
\end{figure}

As a further test, we compute the volume of the 3D ellipsoid for the cosmological parameters as:
\begin{equation}
V=\sqrt{\det{cov(\alpha_{\parallel},\alpha_{\perp},f\sigma_8)}},
\end{equation}
where $cov(\alpha_{\parallel},\alpha_{\perp},f\sigma_8)$ is the parameter covariance that we recover from the full MCMC chains. We show the ratio of $V$ to the N-body result in Figure~\ref{vol_ell}. Deviations larger than $10\%$ are found for \icecola and Lognormal in Sample 1, and for Pinocchio in Sample 2. The single parameter errors shown in Figure~\ref{err_ratio} for these cases have smaller deviations from the N-body result but the off-diagonal elements of the parameter covariance drive the deviations in terms of volume to higher values.

\subsubsection{Dimensionality of parameter space and $k_{max}$}

\begin{figure}
\includegraphics{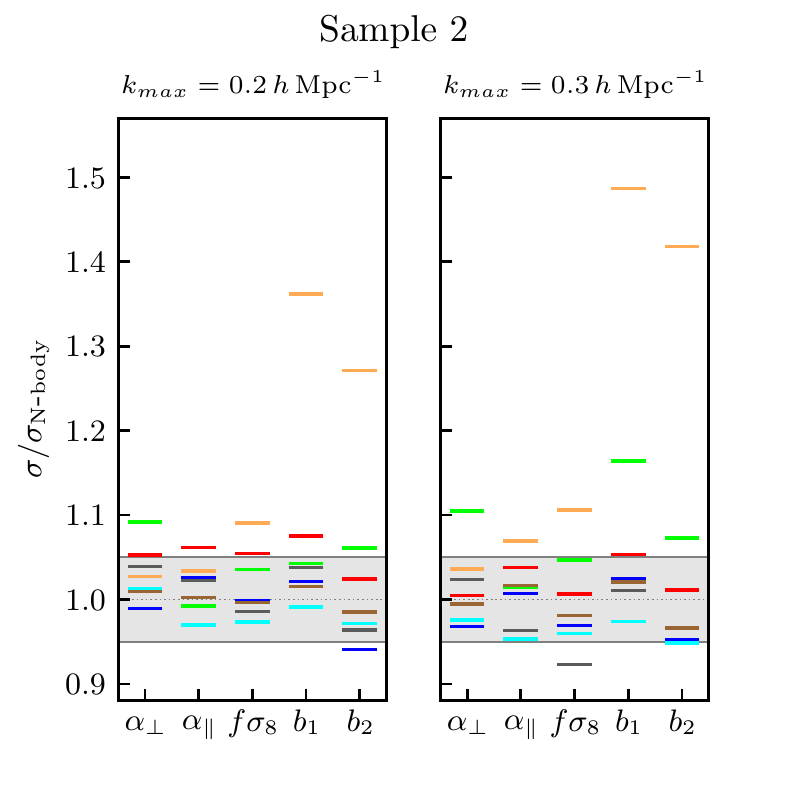}
\caption{{\it Dependence with dimensionality of parameter space and $k_{max}$:} Ratio of the error on cosmological for a reduced set of nuisance parameters ($\gamma_3$ fixed), with respect to the N-body ones, for sample 2. Left panel corresponds to fits with $k_{max}=0.2\kvecMpc$ (our baseline value) and right panel with $k_{max}=0.3\kvecMpc$. The grey shaded area indicate $5\%$ differences. We show the results for \icecola (blue), \pinocchio (red), \peakpatch (brown), \halogen (green), \patchy (cyan), Lognormal (orange) and Gaussian (grey). Our results are stable against these changes, except for Lognormal, whose performance decreases rapidly, and, to a less extent, Halogen.}
\label{err_ratio_kmax}
\end{figure}

We finish by studying the stability of our results against changes of the dimensionality of our parameter space and of the non-linear scale cut-off. This is motivated in part because, despite the bad performance of Lognormal on the monopole covariance, the cosmological parameter errors are reproduced quite well. At the same time the nuisance parameters $b_2$ and $\gamma_3$ are poorly constrained, especially in Sample 2, so that some of the differences between methods can be concealed. In order to test this we run chains by fixing $\gamma_3^-$ to its true value reported in Table~\ref{tab:best_fit} for two different values of $k_{max}$ (one of them our baseline). The results are shown in Figure~\ref{err_ratio_kmax} where we can see that in fact the discrepancy of the error obtained with the Lognormal method increases to $10\%$ for $f\sigma_8$ and can be as large as $50\%$ for the nuisance parameter $b_1$ and $b_2$, with a clear trend of increasing discrepancy with higher $k_{max}$. We also notice that the performances of the other methods are not affected by fixing $\gamma_3^-$ or increasing the $k_{max}$ to $0.3\kvecMpc$, with the exception of Halogen that shows increased ratios on all the parameters and exceeds the $10\%$ level for $\alpha_{\perp}$ and $b_1$ when $k_{max}$ is increased.

%%%%%%%%%%%%%%%%%%%%%%%%%%%%%%%%%%%%%%%%%%%%%%%%
% Conclusions				       %
%%%%%%%%%%%%%%%%%%%%%%%%%%%%%%%%%%%%%%%%%%%%%%%%

\section{Discussion and conclusions}
\label{sec:conclusions}

The cosmological analysis of galaxy redshift surveys requires an estimation of the covariance of the observables to be able to put constraints on cosmological parameters in a Gaussian likelihood analysis framework. One way of doing this is to simulate the galaxy sample of interest a large number of times and computing the covariance using the sample covariance estimator. Since full N-body simulations are computationally very expensive, galaxy mock catalogues used for covariance estimation are produced using so-called ``approximate methods'', that speed up the calculation by introducing approximations in the dynamics or the statistics of the galaxy density field. 

In this work we studied for the first time the accuracy of these methods in reproducing the N-body halo power spectrum multipoles, their covariances and the recovered parameter errors on the set of cosmological and biasing parameters $\{\alpha_{\parallel},\alpha_{\perp},f\sigma_8,b_1,b_2,\gamma^{-}_3\}$. We did this at $z=1$ by comparing ensembles of simulations from approximate methods with a reference set of 300 N-body simulations. To minimise the impact of sample variance errors and noise due to the limited number of simulations, we used the same ICs as the N-body runs for the approximate mocks.

For completeness we considered codes that introduce different types of approximations, trying to span the broad range of algorithms proposed so far in the literature. These can be summarised as follows: fast PM methods, that introduce approximations in the force computation in a PM code, represented by \icecola, Lagrangian methods, that identify halos in Lagrangian space and displace them to the final redshift using LPT, represented by \pinocchio and \peakpatch, bias-based methods that use LPT to generate a matter field at the redshift of interest and applies a biasing scheme to define halos, represented by \patchy and \halogen, and methods that rely on assumptions on the shape of the PDF of the density field, represented by Lognormal mocks and Gaussian covariances. In this order, the grouping transitions from higher resolution and complexity, and hence computational cost, to simpler and faster methods. The methods used in this article, spanning the algorithms above, are presented in Table~\ref{methods_tab}. 

We then compared performances for two halo samples: the first has a linear bias of $\sim 2.6$ and a number density of $\sim2\times10^{-4}\kvecMpccube$ while the second has a linear bias of $\sim 3.4$ and a number density of $\sim5.5\times10^{-5}\kvecMpccube$. Provided with this, we defined halo samples in each approximate method by abundance matching to those two samples above. This was mainly motivated by the fact that different methods define halos in different ways.

Our conclusions can be summarised as follows:

\begin{enumerate}

{\item {\it Power Spectrum Multipole covariance:} The variance of the power spectrum multipoles is in general reproduced within $10\%$ by all methods in both samples, with the exception of the monopole for \halogen and Lognormal. In this regard a special mention is due to Lognormal, and to a less extent \halogen, which show very large variance and covariance for the monopole power spectrum that only reaches the Gaussian expectation on very large scales. For Lognormal this might be due to the fact that our samples are highly biased, in a regime where the lognormal and gaussian fields are not a good approximation to each other. 
Further work is needed to explain this behaviour, which translates into an over-estimation of error bars in the bias parameters.}

{\item {\it Recovered model parameters: } We use the covariance obtained from each method to run a likelihood analysis using a known theory data vector. From the MCMC chains we recover best fit model parameters related to biasing, AP and growth rate of structure and their errors. Although not shown we find that different covariances do not bias the recovery of cosmological parameters, which are always within $1\%$ of their input values, while we find larger (few \%) biases for poorly constrained nuisance parameters.}

{\item  {\it Errors on model parameters:} Reaching conclusions from the translation of differences in the covariances into differences at the level of errors in model parameter $\sigma$ is not trivial because of the finite number of mocks available, which is nonetheless 300 mocks for each method in this work. To partially overcome this limitation we match the initial conditions of the approximated mocks with those of the N-body runs. This reduces the expected error on the ``error ratio'' ($\sigma / \sigma_{\rm N-body}$) by a factor of $\lesssim 2$ to about $4\%-5\%$. Although a rough estimate, this should be regarded as an indication of the statistical limit of our analysis. Bearing this in mind, we find that all the methods recover the error on cosmological parameters for both samples at the $5\%$ level or better, with the exception of \halogen, that has larger deviations for one of the AP parameters in both samples, and Pinocchio in the second sample. The errors on nuisance parameters are within the statistical uncertainty only for \patchy in the first sample and \peakpatch, \icecola, \halogen and Gaussian in the second sample.
In turn, the volume of the 3D ellipsoid defined by the $1\sigma$ contours in the parameter space \{$\alpha_{\parallel} ,\alpha_{\perp} , f\,\sigma_8$\} is reproduced within 10\% by all methods with the exception of \icecola and Lognormal in Sample 1 and \pinocchio in Sample 2. Remarkably the Gaussian prediction (albeit using a nonlinear power spectrum) performs well, presumably because it matches the large-scale bias and the shot-noise independently. In our set up of periodic co-moving boxes the two-point Gaussian term seems dominant against the missing contribution from a four-point connected term. Further exploration of analytical prescriptions is well motivated, e.g. to understand whether the same is true for more realistic situations (see below). In summary: if the requirement for approximate method covariances is to reproduce N-body derived parameter errors at the 5\% level, then this is already achieved by several methods on the most important parameters, but not all. A looser requirement of 10\% is reached by all methods on all parameters, with the only exception of Lognormal.}

{\item Lastly, we find that in general approximate mocks have slightly better performances in the less massive sample (Sample 1) than in Sample 2, especially when looking at the multipoles and their variance. This might be due to a larger non-linear contributions in the higher mass sample, since we observe a similar behaviour in the two point correlation function and bispectrum analysis \citep{paper1,paper3}. There may also be a component of non-Poissonian shot noise from halo exclusion affecting the Fourier space analyses. A detailed analysis of this is beyond the scope of these articles, and we leave it for further work.}

\end{enumerate}

The analysis presented in this series of articles can be regarded as of general applicability for the next generation of galaxy surveys, but it has been developed with particular attention to scientific requirements of the {\it Euclid} mission. Despite of the simplified setting used in this first analysis, these results confirm that usage of approximate methods for estimating the
covariance matrix of power spectra is allowed at the cost of a small systematic error on the recovered parameter uncertainty (generally within $10\%$, i.e. complying with {\it Euclid} requirements).
Even though great care was put into minimizing the impact of noise in the comparison, the strength of our conclusions are somewhat limited by the number of mocks that we are using. 
We thus plan to perform follow-up studies with larger number of approximate mocks and to study more interesting regimes of mass resolution, that would allow us to reach number densities that are closer to those in future galaxy surveys. In addition, we have considered a rather simplistic scenario of halo snapshots in comoving outputs with periodic boundary conditions. Further work will incorporate masking effects, addition of galaxies into halos (e.g. in particular halos populated using an HOD prescription for galaxies matching the $H_\alpha$ emitters that {\it Euclid} will observe) and inclusion of super-survey modes. We expect that the inclusion of these effects will enhance covariances through the coupling between long and short modes, possibly making differences among methods more evident. Moreover, the modelling of such effects is not trivial in an analytical framework, such as the Gaussian approximation presented in this article, thus it will be interesting to test the performance of this method in more realistic scenarios.

%%%%%%%%%%%%%%%%%%%%%%%%%%%%%%%%%%%%%%%%%%%%%%%%
% acknowledgments			       %
%%%%%%%%%%%%%%%%%%%%%%%%%%%%%%%%%%%%%%%%%%%%%%%%

\section*{Acknowledgments}

This article and companion articles have benefited of discussions and the stimulating 
environment of the Euclid Consortium, which is warmly acknowledged. 

L. Blot acknowledges support from the Spanish Ministerio de Econom\'ia y Competitividad (MINECO) grant ESP2015-66861. M.Crocce acknowledges support from the Spanish Ram\'on y Cajal MICINN program. M.Crocce has been funded by AYA2015-71825.

M. Colavincenzo is supported by the {\em Departments of Excellence 2018 - 2022} Grant awarded by the Italian Ministero dell'Istruzione, dell'Universit\`a e della Ricerca (MIUR) (L. 232/2016), by the research grant {\em The Anisotropic Dark Universe} Number CSTO161409, funded under the program CSP-UNITO {\em Research for the Territory 2016} by Compagnia di Sanpaolo and University of Torino; and the research grant TAsP (Theoretical Astroparticle Physics) funded by the Istituto Nazionale di Fisica Nucleare (INFN). P. Monaco and E. Sefusatti acknowledge support from grant MIUR PRIN 2015 {\em Cosmology and Fundamental Physics: illuminating the Dark Universe with Euclid} and from Consorzio per la Fisica di Trieste; they are part of the INFN InDark research group.

M. Lippich and A.G.S\'anchez acknowledge support from the Transregional Collaborative Research Centre TR33 {\em The Dark Universe} of the German Research Foundation (DFG). 

C. Dalla Vecchia acknowledges support from the MINECO through grants AYA2013-46886, AYA2014-58308 and RYC-2015-18078. S. Avila acknowledges support from the UK Space Agency through grant ST/K00283X/1. A. Balaguera-Antol\'{i}nez acknowledges financial support from MINECO under the Severo Ochoa program SEV-2015-0548. M. Pellejero-Ibanez acknowledges support from MINECO under the grand AYA2012-39702-C02-01. P. Fosalba acknowledges support from MINECO through grant ESP2015-66861-C3-1-R and Generalitat de Catalunya through grant 2017-SGR-885. A. Izard was supported in part by Jet Propulsion Laboratory, California Institute of Technology, under a contract with the National Aeronautics and Space Administration. He was also supported in part by NASA ROSES 13-ATP13-0019, NASA ROSES 14-MIRO-PROs-0064, NASA ROSES 12- EUCLID12-0004, and acknowledges support from the JAE program grant from the Spanish National Science Council (CSIC). R. Bond, S. Codis and G. Stein are supported by the Canadian Natural Sciences and Engineering Research Council (NSERC). G. Yepes acknowledges financial support from MINECO/FEDER (Spain) under research grant AYA2015-63810-P. 

The Minerva simulations have been performed and analysed on the Hydra cluster and on the computing cluster for the \textit{Euclid} project at the Max Planck Computing and Data Facility (MPCDF) in Garching.

\icecola simulations were run at the MareNostrum supercomputer - Barcelona Supercomputing Center (BSC-CNS, www.bsc.es), through the grant AECT-2016- 3-0015. 

\pinocchio mocks were run on the GALILEO cluster at CINECA thanks to an agreement with  the University of Trieste.

\peakpatch simulations were performed on the GPC supercomputer at the SciNet HPC Consortium. SciNet is funded by: the Canada Foundation for Innovation under the auspices of Compute Canada; the Government of Ontario; Ontario Research Fund - Research Excellence; and the University of Toronto.

Numerical computations with \halogen were done on the Sciama High Performance Compute (HPC) cluster which is supported by the ICG, SEPNet and the University of Portsmouth.

\patchy mocks have been computed in part at the MareNostrum supercomputer of the Barcelona Supercomputing Center thanks to a grant from the Red Espa\~nola de Supercomputaci\'on (RES), and in part at the Teide High-Performance Computing facilities provided by the Instituto Tecnol\'ogico y de Energ\'{\i}as Renovables (ITER, S.A.).

This work has made use of the NumPy, Matplotlib and ChainConsumer Python packages. 

\bibliographystyle{mnras}
\bibliography{cosmologia}

\begin{thebibliography}{}
\makeatletter
\relax
\def\mn@urlcharsother{\let\do\@makeother \do\$\do\&\do\#\do\^\do\_\do\%\do\~}
\def\mn@doi{\begingroup\mn@urlcharsother \@ifnextchar [ {\mn@doi@}
  {\mn@doi@[]}}
\def\mn@doi@[#1]#2{\def\@tempa{#1}\ifx\@tempa\@empty \href
  {http://dx.doi.org/#2} {doi:#2}\else \href {http://dx.doi.org/#2} {#1}\fi
  \endgroup}
\def\mn@eprint#1#2{\mn@eprint@#1:#2::\@nil}
\def\mn@eprint@arXiv#1{\href {http://arxiv.org/abs/#1} {{\tt arXiv:#1}}}
\def\mn@eprint@dblp#1{\href {http://dblp.uni-trier.de/rec/bibtex/#1.xml}
  {dblp:#1}}
\def\mn@eprint@#1:#2:#3:#4\@nil{\def\@tempa {#1}\def\@tempb {#2}\def\@tempc
  {#3}\ifx \@tempc \@empty \let \@tempc \@tempb \let \@tempb \@tempa \fi \ifx
  \@tempb \@empty \def\@tempb {arXiv}\fi \@ifundefined
  {mn@eprint@\@tempb}{\@tempb:\@tempc}{\expandafter \expandafter \csname
  mn@eprint@\@tempb\endcsname \expandafter{\@tempc}}}

\bibitem[\protect\citeauthoryear{{Agrawal}, {Makiya}, {Chiang}, {Jeong},
  {Saito}  \& {Komatsu}}{{Agrawal} et~al.}{2017}]{AgrawalEtal2017}
{Agrawal} A.,  {Makiya} R.,  {Chiang} C.-T.,  {Jeong} D.,  {Saito} S.,
  {Komatsu} E.,  2017, \mn@doi [\jcap] {10.1088/1475-7516/2017/10/003}, \href
  {http://adsabs.harvard.edu/abs/2017JCAP...10..003A} {10, 003}

\bibitem[\protect\citeauthoryear{{Alam} et~al.,}{{Alam}
  et~al.}{2017}]{2017MNRAS.470.2617A}
{Alam} S.,  et~al., 2017, \mn@doi [\mnras] {10.1093/mnras/stx721}, \href
  {http://adsabs.harvard.edu/abs/2017MNRAS.470.2617A} {470, 2617}

\bibitem[\protect\citeauthoryear{{Alcock} \& {Paczynski}}{{Alcock} \&
  {Paczynski}}{1979}]{AP}
{Alcock} C.,  {Paczynski} B.,  1979, \mn@doi [\nat] {10.1038/281358a0}, \href
  {http://adsabs.harvard.edu/abs/1979Natur.281..358A} {281, 358}

\bibitem[\protect\citeauthoryear{{Alvarez} et~al.}{{Alvarez}
  et~al.}{2018}]{Alvarezinprep}
{Alvarez} M.,  et~al., 2018, in prep.

\bibitem[\protect\citeauthoryear{{Angulo}, {Springel}, {White}, {Jenkins},
  {Baugh}  \& {Frenk}}{{Angulo} et~al.}{2012}]{2012MNRAS.426.2046A}
{Angulo} R.~E.,  {Springel} V.,  {White} S.~D.~M.,  {Jenkins} A.,  {Baugh}
  C.~M.,   {Frenk} C.~S.,  2012, \mn@doi [\mnras]
  {10.1111/j.1365-2966.2012.21830.x}, \href
  {http://adsabs.harvard.edu/abs/2012MNRAS.426.2046A} {426, 2046}

\bibitem[\protect\citeauthoryear{{Ata} et~al.,}{{Ata}
  et~al.}{2018}]{2018MNRAS.473.4773A}
{Ata} M.,  et~al., 2018, \mn@doi [\mnras] {10.1093/mnras/stx2630}, \href
  {http://adsabs.harvard.edu/abs/2018MNRAS.473.4773A} {473, 4773}

\bibitem[\protect\citeauthoryear{{Avila}, {Murray}, {Knebe}, {Power},
  {Robotham}  \& {Garcia-Bellido}}{{Avila} et~al.}{2015}]{AvilaEtal2015}
{Avila} S.,  {Murray} S.~G.,  {Knebe} A.,  {Power} C.,  {Robotham} A.~S.~G.,
  {Garcia-Bellido} J.,  2015, \mn@doi [\mnras] {10.1093/mnras/stv711}, \href
  {http://adsabs.harvard.edu/abs/2015MNRAS.450.1856A} {450, 1856}

\bibitem[\protect\citeauthoryear{{Avila} et~al.,}{{Avila}
  et~al.}{2017}]{Avila2017}
{Avila} S.,  et~al., 2017, preprint, \href
  {http://adsabs.harvard.edu/abs/2017arXiv171206232A} {} (\mn@eprint {arXiv}
  {1712.06232})

\bibitem[\protect\citeauthoryear{{Ballinger}, {Peacock}  \&
  {Heavens}}{{Ballinger} et~al.}{1996}]{1996MNRAS.282..877B}
{Ballinger} W.~E.,  {Peacock} J.~A.,   {Heavens} A.~F.,  1996, \mn@doi [\mnras]
  {10.1093/mnras/282.3.877}, \href
  {http://adsabs.harvard.edu/abs/1996MNRAS.282..877B} {282, 877}

\bibitem[\protect\citeauthoryear{{Bond} \& {Myers}}{{Bond} \&
  {Myers}}{1996a}]{BondMyers1996A}
{Bond} J.~R.,  {Myers} S.~T.,  1996a, \mn@doi [\apjs] {10.1086/192267}, \href
  {http://adsabs.harvard.edu/abs/1996ApJS..103....1B} {103, 1}

\bibitem[\protect\citeauthoryear{{Bond} \& {Myers}}{{Bond} \&
  {Myers}}{1996b}]{BondMyers1996B}
{Bond} J.~R.,  {Myers} S.~T.,  1996b, \mn@doi [\apjs] {10.1086/192268}, \href
  {http://adsabs.harvard.edu/abs/1996ApJS..103...41B} {103, 41}

\bibitem[\protect\citeauthoryear{{Bond} \& {Myers}}{{Bond} \&
  {Myers}}{1996c}]{BondMyers1996C}
{Bond} J.~R.,  {Myers} S.~T.,  1996c, \mn@doi [\apjs] {10.1086/192269}, \href
  {http://adsabs.harvard.edu/abs/1996ApJS..103...63B} {103, 63}

\bibitem[\protect\citeauthoryear{{Carretero}, {Castander}, {Gazta{\~n}aga},
  {Crocce}  \& {Fosalba}}{{Carretero} et~al.}{2015}]{2015MNRAS.447..646C}
{Carretero} J.,  {Castander} F.~J.,  {Gazta{\~n}aga} E.,  {Crocce} M.,
  {Fosalba} P.,  2015, \mn@doi [\mnras] {10.1093/mnras/stu2402}, \href
  {http://adsabs.harvard.edu/abs/2015MNRAS.447..646C} {447, 646}

\bibitem[\protect\citeauthoryear{{Chan}, {Scoccimarro}  \& {Sheth}}{{Chan}
  et~al.}{2012}]{2012PhRvD..85h3509C}
{Chan} K.~C.,  {Scoccimarro} R.,   {Sheth} R.~K.,  2012, \mn@doi [\prd]
  {10.1103/PhysRevD.85.083509}, \href
  {http://adsabs.harvard.edu/abs/2012PhRvD..85h3509C} {85, 083509}

\bibitem[\protect\citeauthoryear{{Chuang}, {Kitaura}, {Prada}, {Zhao}  \&
  {Yepes}}{{Chuang} et~al.}{2015a}]{2015MNRAS.446.2621C}
{Chuang} C.-H.,  {Kitaura} F.-S.,  {Prada} F.,  {Zhao} C.,   {Yepes} G.,
  2015a, \mn@doi [\mnras] {10.1093/mnras/stu2301}, \href
  {http://adsabs.harvard.edu/abs/2015MNRAS.446.2621C} {446, 2621}

\bibitem[\protect\citeauthoryear{{Chuang} et~al.,}{{Chuang}
  et~al.}{2015b}]{2015MNRAS.452..686C}
{Chuang} C.-H.,  et~al., 2015b, \mn@doi [\mnras] {10.1093/mnras/stv1289}, \href
  {http://adsabs.harvard.edu/abs/2015MNRAS.452..686C} {452, 686}

\bibitem[\protect\citeauthoryear{{Colavincenzo} et~al.,}{{Colavincenzo}
  et~al.}{2019}]{paper3}
{Colavincenzo} M.,  et~al., 2019, \mn@doi [\mnras] {10.1093/mnras/sty2964},
  \href {http://adsabs.harvard.edu/abs/2019MNRAS.482.4883C} {482, 4883}

\bibitem[\protect\citeauthoryear{{Cole} et~al.,}{{Cole}
  et~al.}{2005}]{2005MNRAS.362..505C}
{Cole} S.,  et~al., 2005, \mn@doi [\mnras] {10.1111/j.1365-2966.2005.09318.x},
  \href {http://adsabs.harvard.edu/abs/2005MNRAS.362..505C} {362, 505}

\bibitem[\protect\citeauthoryear{{DESI Collaboration} et~al.,}{{DESI
  Collaboration} et~al.}{2016}]{2016arXiv161100036D}
{DESI Collaboration} et~al., 2016, preprint, \href
  {http://adsabs.harvard.edu/abs/2016arXiv161100036D} {} (\mn@eprint {arXiv}
  {1611.00036})

\bibitem[\protect\citeauthoryear{{Dawson} et~al.,}{{Dawson}
  et~al.}{2016}]{2016AJ....151...44D}
{Dawson} K.~S.,  et~al., 2016, \mn@doi [\aj] {10.3847/0004-6256/151/2/44},
  \href {http://adsabs.harvard.edu/abs/2016AJ....151...44D} {151, 44}

\bibitem[\protect\citeauthoryear{{Dodelson} \& {Schneider}}{{Dodelson} \&
  {Schneider}}{2013}]{DodelsonSchneider2013}
{Dodelson} S.,  {Schneider} M.~D.,  2013, \mn@doi [\prd]
  {10.1103/PhysRevD.88.063537}, \href
  {http://adsabs.harvard.edu/abs/2013PhRvD..88f3537D} {88, 063537}

\bibitem[\protect\citeauthoryear{{Eisenstein} et~al.,}{{Eisenstein}
  et~al.}{2005}]{2005ApJ...633..560E}
{Eisenstein} D.~J.,  et~al., 2005, \mn@doi [\apj] {10.1086/466512}, \href
  {http://adsabs.harvard.edu/abs/2005ApJ...633..560E} {633, 560}

\bibitem[\protect\citeauthoryear{{Feng}, {Chu}, {Seljak}  \& {McDonald}}{{Feng}
  et~al.}{2016}]{2016MNRAS.463.2273F}
{Feng} Y.,  {Chu} M.-Y.,  {Seljak} U.,   {McDonald} P.,  2016, \mn@doi [\mnras]
  {10.1093/mnras/stw2123}, \href
  {http://adsabs.harvard.edu/abs/2016MNRAS.463.2273F} {463, 2273}

\bibitem[\protect\citeauthoryear{{Fosalba}, {Crocce}, {Gazta{\~n}aga}  \&
  {Castander}}{{Fosalba} et~al.}{2015}]{2015MNRAS.448.2987F}
{Fosalba} P.,  {Crocce} M.,  {Gazta{\~n}aga} E.,   {Castander} F.~J.,  2015,
  \mn@doi [\mnras] {10.1093/mnras/stv138}, \href
  {http://adsabs.harvard.edu/abs/2015MNRAS.448.2987F} {448, 2987}

\bibitem[\protect\citeauthoryear{Grieb, S{\'a}nchez, Salazar-Albornoz  \&
  Dalla~Vecchia}{Grieb et~al.}{2016}]{GriebEtal2016}
Grieb J.~N.,  S{\'a}nchez A.~G.,  Salazar-Albornoz S.,   Dalla~Vecchia C.,
  2016, \mn@doi [\mnras] {10.1093/mnras/stw065}, \href
  {http://adsabs.harvard.edu/abs/2016MNRAS.457.1577G} {457, 1577}

\bibitem[\protect\citeauthoryear{{Grieb} et~al.,}{{Grieb}
  et~al.}{2017}]{2017MNRAS.467.2085G}
{Grieb} J.~N.,  et~al., 2017, \mn@doi [\mnras] {10.1093/mnras/stw3384}, \href
  {http://adsabs.harvard.edu/abs/2017MNRAS.467.2085G} {467, 2085}

\bibitem[\protect\citeauthoryear{{Hartlap}, {Simon}  \& {Schneider}}{{Hartlap}
  et~al.}{2007}]{HartlapSimonSchneider2007}
{Hartlap} J.,  {Simon} P.,   {Schneider} P.,  2007, \mn@doi [\aap]
  {10.1051/0004-6361:20066170}, \href
  {http://adsabs.harvard.edu/abs/2007A%26A...464..399H} {464, 399}

\bibitem[\protect\citeauthoryear{{Heitmann} et~al.,}{{Heitmann}
  et~al.}{2015}]{2015ApJS..219...34H}
{Heitmann} K.,  et~al., 2015, \mn@doi [\apjs] {10.1088/0067-0049/219/2/34},
  \href {http://adsabs.harvard.edu/abs/2015ApJS..219...34H} {219, 34}

\bibitem[\protect\citeauthoryear{{Hildebrandt} et~al.,}{{Hildebrandt}
  et~al.}{2017}]{2017MNRAS.465.1454H}
{Hildebrandt} H.,  et~al., 2017, \mn@doi [\mnras] {10.1093/mnras/stw2805},
  \href {http://adsabs.harvard.edu/abs/2017MNRAS.465.1454H} {465, 1454}

\bibitem[\protect\citeauthoryear{{Howlett}, {Ross}, {Samushia}, {Percival}  \&
  {Manera}}{{Howlett} et~al.}{2015}]{Howlett2015}
{Howlett} C.,  {Ross} A.~J.,  {Samushia} L.,  {Percival} W.~J.,   {Manera} M.,
  2015, \mn@doi [\mnras] {10.1093/mnras/stu2693}, \href
  {http://adsabs.harvard.edu/abs/2015MNRAS.449..848H} {449, 848}

\bibitem[\protect\citeauthoryear{{Ivezic} et~al.,}{{Ivezic}
  et~al.}{2008}]{2008arXiv0805.2366I}
{Ivezic} Z.,  et~al., 2008, preprint, \href
  {http://adsabs.harvard.edu/abs/2008arXiv0805.2366I} {} (\mn@eprint {arXiv}
  {0805.2366})

\bibitem[\protect\citeauthoryear{{Izard}, {Crocce}  \& {Fosalba}}{{Izard}
  et~al.}{2016}]{IzardCrocceFosalba2016}
{Izard} A.,  {Crocce} M.,   {Fosalba} P.,  2016, \mn@doi [\mnras]
  {10.1093/mnras/stw797}, \href
  {http://adsabs.harvard.edu/abs/2016MNRAS.459.2327I} {459, 2327}

\bibitem[\protect\citeauthoryear{{Izard}, {Fosalba}  \& {Crocce}}{{Izard}
  et~al.}{2018}]{2015arXiv150904685I}
{Izard} A.,  {Fosalba} P.,   {Crocce} M.,  2018, \mn@doi [\mnras]
  {10.1093/mnras/stx2544}, \href
  {http://adsabs.harvard.edu/abs/2018MNRAS.473.3051I} {473, 3051}

\bibitem[\protect\citeauthoryear{{Kitaura}, {Yepes}  \& {Prada}}{{Kitaura}
  et~al.}{2014}]{Kitaura2014}
{Kitaura} F.-S.,  {Yepes} G.,   {Prada} F.,  2014, \mn@doi [\mnras]
  {10.1093/mnrasl/slt172}, \href
  {http://adsabs.harvard.edu/abs/2014MNRAS.439L..21K} {439, L21}

\bibitem[\protect\citeauthoryear{{Kitaura}, {Gil-Mar{\'{\i}}n}, {Sc{\'o}ccola},
  {Chuang}, {M{\"u}ller}, {Yepes}  \& {Prada}}{{Kitaura}
  et~al.}{2015}]{KitauraEtal2015}
{Kitaura} F.-S.,  {Gil-Mar{\'{\i}}n} H.,  {Sc{\'o}ccola} C.~G.,  {Chuang}
  C.-H.,  {M{\"u}ller} V.,  {Yepes} G.,   {Prada} F.,  2015, \mn@doi [\mnras]
  {10.1093/mnras/stv645}, \href
  {http://adsabs.harvard.edu/abs/2015MNRAS.450.1836K} {450, 1836}

\bibitem[\protect\citeauthoryear{{Kitaura} et~al.,}{{Kitaura}
  et~al.}{2016}]{Kitaura2016}
{Kitaura} F.-S.,  et~al., 2016, \mn@doi [\mnras] {10.1093/mnras/stv2826}, \href
  {http://adsabs.harvard.edu/abs/2016MNRAS.456.4156K} {456, 4156}

\bibitem[\protect\citeauthoryear{{Koda}, {Blake}, {Beutler}, {Kazin}  \&
  {Marin}}{{Koda} et~al.}{2016}]{Koda2016}
{Koda} J.,  {Blake} C.,  {Beutler} F.,  {Kazin} E.,   {Marin} F.,  2016,
  \mn@doi [\mnras] {10.1093/mnras/stw763}, \href
  {http://adsabs.harvard.edu/abs/2016MNRAS.459.2118K} {459, 2118}

\bibitem[\protect\citeauthoryear{{LSST Science Collaboration} et~al.,}{{LSST
  Science Collaboration} et~al.}{2009}]{2009arXiv0912.0201L}
{LSST Science Collaboration} et~al., 2009, preprint, \href
  {http://adsabs.harvard.edu/abs/2009arXiv0912.0201L} {} (\mn@eprint {arXiv}
  {0912.0201})

\bibitem[\protect\citeauthoryear{{Laureijs} et~al.,}{{Laureijs}
  et~al.}{2011}]{LaureijsEtal2011}
{Laureijs} R.,  et~al., 2011, ArXiv: 1110.3193, \href
  {http://adsabs.harvard.edu/abs/2011arXiv1110.3193L} {}

\bibitem[\protect\citeauthoryear{{Lippich} et~al.,}{{Lippich}
  et~al.}{2019}]{paper1}
{Lippich} M.,  et~al., 2019, \mn@doi [\mnras] {10.1093/mnras/sty2757}, \href
  {http://adsabs.harvard.edu/abs/2019MNRAS.482.1786L} {482, 1786}

\bibitem[\protect\citeauthoryear{{Mandelbaum} et~al.,}{{Mandelbaum}
  et~al.}{2018}]{2018PASJ...70S..25M}
{Mandelbaum} R.,  et~al., 2018, \mn@doi [\pasj] {10.1093/pasj/psx130}, \href
  {http://adsabs.harvard.edu/abs/2018PASJ...70S..25M} {70, S25}

\bibitem[\protect\citeauthoryear{Manera et~al.,}{Manera
  et~al.}{2013}]{Maneraetal2013}
Manera M.,  et~al., 2013, \mn@doi [\mnras] {10.1093/mnras/sts084}, \href
  {http://adsabs.harvard.edu/abs/2013MNRAS.428.1036M} {428, 1036}

\bibitem[\protect\citeauthoryear{{Manera} et~al.,}{{Manera}
  et~al.}{2015}]{Maneraetal2015}
{Manera} M.,  et~al., 2015, \mn@doi [\mnras] {10.1093/mnras/stu2465}, \href
  {http://adsabs.harvard.edu/abs/2015MNRAS.447..437M} {447, 437}

\bibitem[\protect\citeauthoryear{{McDonald}}{{McDonald}}{2006}]{2006PhRvD..74j3512M}
{McDonald} P.,  2006, \mn@doi [\prd] {10.1103/PhysRevD.74.103512}, \href
  {https://ui.adsabs.harvard.edu/abs/2006PhRvD..74j3512M} {74, 103512}

\bibitem[\protect\citeauthoryear{{Monaco}}{{Monaco}}{2016}]{monaco2016}
{Monaco} P.,  2016, \mn@doi [Galaxies] {10.3390/galaxies4040053}, \href
  {http://adsabs.harvard.edu/abs/2016Galax...4...53M} {4, 53}

\bibitem[\protect\citeauthoryear{Monaco, {Theuns}, {Taffoni}, {Governato},
  {Quinn}  \& {Stadel}}{Monaco et~al.}{2002}]{MonacoEtal2002}
Monaco P.,  {Theuns} T.,  {Taffoni} G.,  {Governato} F.,  {Quinn} T.,
  {Stadel} J.,  2002, \mn@doi [\apj] {10.1086/324182}, \href
  {http://adsabs.harvard.edu/abs/2002ApJ...564....8M} {564, 8}

\bibitem[\protect\citeauthoryear{Monaco, Sefusatti, Borgani, Crocce, Fosalba,
  Sheth  \& Theuns}{Monaco et~al.}{2013}]{MonacoEtal2013}
Monaco P.,  Sefusatti E.,  Borgani S.,  Crocce M.,  Fosalba P.,  Sheth R.~K.,
  Theuns T.,  2013, \mn@doi [\mnras] {10.1093/mnras/stt907}, \href
  {http://adsabs.harvard.edu/abs/2013MNRAS.433.2389M} {433, 2389}

\bibitem[\protect\citeauthoryear{{Munari}, {Monaco}, {Koda}, {Kitaura},
  {Sefusatti}  \& {Borgani}}{{Munari} et~al.}{2017a}]{MunariEtal2017B}
{Munari} E.,  {Monaco} P.,  {Koda} J.,  {Kitaura} F.-S.,  {Sefusatti} E.,
  {Borgani} S.,  2017a, \mn@doi [\jcap] {10.1088/1475-7516/2017/07/050}, \href
  {http://adsabs.harvard.edu/abs/2017JCAP...07..050M} {7, 050}

\bibitem[\protect\citeauthoryear{{Munari}, {Monaco}, {Sefusatti}, {Castorina},
  {Mohammad}, {Anselmi}  \& {Borgani}}{{Munari} et~al.}{2017b}]{MunariEtal2017}
{Munari} E.,  {Monaco} P.,  {Sefusatti} E.,  {Castorina} E.,  {Mohammad} F.~G.,
   {Anselmi} S.,   {Borgani} S.,  2017b, \mn@doi [\mnras]
  {10.1093/mnras/stw3085}, \href
  {http://adsabs.harvard.edu/abs/2017MNRAS.465.4658M} {465, 4658}

\bibitem[\protect\citeauthoryear{{Nishimichi} \& {Taruya}}{{Nishimichi} \&
  {Taruya}}{2011}]{eTNS}
{Nishimichi} T.,  {Taruya} A.,  2011, \mn@doi [\prd]
  {10.1103/PhysRevD.84.043526}, \href
  {http://adsabs.harvard.edu/abs/2011PhRvD..84d3526N} {84, 043526}

\bibitem[\protect\citeauthoryear{{Percival} et~al.,}{{Percival}
  et~al.}{2001}]{PercivalEtal2001}
{Percival} W.~J.,  et~al., 2001, \mn@doi [\mnras]
  {10.1046/j.1365-8711.2001.04827.x}, \href
  {http://adsabs.harvard.edu/abs/2001MNRAS.327.1297P} {327, 1297}

\bibitem[\protect\citeauthoryear{{Percival} et~al.,}{{Percival}
  et~al.}{2014}]{PercivalEtal2014}
{Percival} W.~J.,  et~al., 2014, \mn@doi [\mnras] {10.1093/mnras/stu112}, \href
  {http://adsabs.harvard.edu/abs/2014MNRAS.439.2531P} {439, 2531}

\bibitem[\protect\citeauthoryear{{Pezzotta} et~al.,}{{Pezzotta}
  et~al.}{2017}]{2017A&A...604A..33P}
{Pezzotta} A.,  et~al., 2017, \mn@doi [\aap] {10.1051/0004-6361/201630295},
  \href {http://adsabs.harvard.edu/abs/2017A\%26A...604A..33P} {604, A33}

\bibitem[\protect\citeauthoryear{{Potter}, {Stadel}  \& {Teyssier}}{{Potter}
  et~al.}{2017}]{2017ComAC...4....2P}
{Potter} D.,  {Stadel} J.,   {Teyssier} R.,  2017, \mn@doi [Computational
  Astrophysics and Cosmology] {10.1186/s40668-017-0021-1}, \href
  {http://adsabs.harvard.edu/abs/2017ComAC...4....2P} {4, 2}

\bibitem[\protect\citeauthoryear{{Samushia}, {Percival}  \&
  {Raccanelli}}{{Samushia} et~al.}{2012}]{2012MNRAS.420.2102S}
{Samushia} L.,  {Percival} W.~J.,   {Raccanelli} A.,  2012, \mn@doi [\mnras]
  {10.1111/j.1365-2966.2011.20169.x}, \href
  {http://adsabs.harvard.edu/abs/2012MNRAS.420.2102S} {420, 2102}

\bibitem[\protect\citeauthoryear{{S{\'a}nchez} et~al.,}{{S{\'a}nchez}
  et~al.}{2013}]{SanchezEtal2013}
{S{\'a}nchez} A.~G.,  et~al., 2013, \mn@doi [\mnras] {10.1093/mnras/stt799},
  \href {http://adsabs.harvard.edu/abs/2013MNRAS.433.1202S} {433, 1202}

\bibitem[\protect\citeauthoryear{{S{\'a}nchez} et~al.,}{{S{\'a}nchez}
  et~al.}{2017}]{SanchezEtal2017b}
{S{\'a}nchez} A.~G.,  et~al., 2017, \mn@doi [\mnras] {10.1093/mnras/stw2443},
  \href {http://adsabs.harvard.edu/abs/2017MNRAS.464.1640S} {464, 1640}

\bibitem[\protect\citeauthoryear{Scoccimarro, Zaldarriaga  \& Hui}{Scoccimarro
  et~al.}{1999}]{ScoccimarroZaldarriagaHui1999}
Scoccimarro R.,  Zaldarriaga M.,   Hui L.,  1999, \mn@doi [\apj]
  {10.1086/308059}, \href {http://adsabs.harvard.edu/abs/1999ApJ...527....1S}
  {527, 1}

\bibitem[\protect\citeauthoryear{{Sefusatti}, {Crocce}, {Scoccimarro}  \&
  {Couchman}}{{Sefusatti} et~al.}{2016}]{SefusattiEtal2016}
{Sefusatti} E.,  {Crocce} M.,  {Scoccimarro} R.,   {Couchman} H.~M.~P.,  2016,
  \mn@doi [\mnras] {10.1093/mnras/stw1229}, \href
  {http://adsabs.harvard.edu/abs/2016MNRAS.460.3624S} {460, 3624}

\bibitem[\protect\citeauthoryear{{Smith}, {Cole}, {Baugh}, {Zheng}, {Angulo},
  {Norberg}  \& {Zehavi}}{{Smith} et~al.}{2017}]{2017MNRAS.470.4646S}
{Smith} A.,  {Cole} S.,  {Baugh} C.,  {Zheng} Z.,  {Angulo} R.,  {Norberg} P.,
   {Zehavi} I.,  2017, \mn@doi [\mnras] {10.1093/mnras/stx1432}, \href
  {http://adsabs.harvard.edu/abs/2017MNRAS.470.4646S} {470, 4646}

\bibitem[\protect\citeauthoryear{{Springel}}{{Springel}}{2005}]{Springel2005}
{Springel} V.,  2005, \mn@doi [\mnras] {10.1111/j.1365-2966.2005.09655.x},
  \href {http://adsabs.harvard.edu/abs/2005MNRAS.364.1105S} {364, 1105}

\bibitem[\protect\citeauthoryear{{Springel}, {White}, {Tormen}  \&
  {Kauffmann}}{{Springel} et~al.}{2001}]{SpringelEtal2001}
{Springel} V.,  {White} S.~D.~M.,  {Tormen} G.,   {Kauffmann} G.,  2001,
  \mn@doi [\mnras] {10.1046/j.1365-8711.2001.04912.x}, \href
  {http://adsabs.harvard.edu/abs/2001MNRAS.328..726S} {328, 726}

\bibitem[\protect\citeauthoryear{{Stein}, {Alvarez}  \& {Bond}}{{Stein}
  et~al.}{2018}]{Steinetal2018}
{Stein} G.,  {Alvarez} M.~A.,   {Bond} J.~R.,  2018, preprint (\mn@eprint
  {arXiv} {1810.07727})

\bibitem[\protect\citeauthoryear{{Takahashi}, {Hamana}, {Shirasaki},
  {Namikawa}, {Nishimichi}, {Osato}  \& {Shiroyama}}{{Takahashi}
  et~al.}{2017}]{2017ApJ...850...24T}
{Takahashi} R.,  {Hamana} T.,  {Shirasaki} M.,  {Namikawa} T.,  {Nishimichi}
  T.,  {Osato} K.,   {Shiroyama} K.,  2017, \mn@doi [\apj]
  {10.3847/1538-4357/aa943d}, \href
  {http://adsabs.harvard.edu/abs/2017ApJ...850...24T} {850, 24}

\bibitem[\protect\citeauthoryear{{Taruya}, {Nishimichi}  \& {Saito}}{{Taruya}
  et~al.}{2010}]{TNS}
{Taruya} A.,  {Nishimichi} T.,   {Saito} S.,  2010, \mn@doi [\prd]
  {10.1103/PhysRevD.82.063522}, \href
  {http://adsabs.harvard.edu/abs/2010PhRvD..82f3522T} {82, 063522}

\bibitem[\protect\citeauthoryear{{Tassev}, {Zaldarriaga}  \&
  {Eisenstein}}{{Tassev} et~al.}{2013}]{TassevZaldarriagaEisenstein2013}
{Tassev} S.,  {Zaldarriaga} M.,   {Eisenstein} D.~J.,  2013, \mn@doi [\jcap]
  {10.1088/1475-7516/2013/06/036}, \href
  {http://adsabs.harvard.edu/abs/2013JCAP...06..036T} {6, 036}

\bibitem[\protect\citeauthoryear{Taylor, Joachimi  \& {Kitching}}{Taylor
  et~al.}{2013}]{TaylorJoachimiKitching2013}
Taylor A.~N.,  Joachimi B.,   {Kitching} T.~D.,  2013, \mn@doi [\mnras]
  {10.1093/mnras/stt270}, \href
  {http://adsabs.harvard.edu/abs/2013MNRAS.432.1928T} {432, 1928}

\bibitem[\protect\citeauthoryear{{Tegmark} et~al.,}{{Tegmark}
  et~al.}{2004}]{2004PhRvD..69j3501T}
{Tegmark} M.,  et~al., 2004, \mn@doi [\prd] {10.1103/PhysRevD.69.103501}, \href
  {http://adsabs.harvard.edu/abs/2004PhRvD..69j3501T} {69, 103501}

\bibitem[\protect\citeauthoryear{{Troxel} et~al.,}{{Troxel}
  et~al.}{2017}]{2017arXiv170801538T}
{Troxel} M.~A.,  et~al., 2017, preprint, \href
  {http://adsabs.harvard.edu/abs/2017arXiv170801538T} {} (\mn@eprint {arXiv}
  {1708.01538})

\bibitem[\protect\citeauthoryear{{White}, {Tinker}  \& {McBride}}{{White}
  et~al.}{2014}]{2014MNRAS.437.2594W}
{White} M.,  {Tinker} J.~L.,   {McBride} C.~K.,  2014, \mn@doi [\mnras]
  {10.1093/mnras/stt2071}, \href
  {http://adsabs.harvard.edu/abs/2014MNRAS.437.2594W} {437, 2594}

\bibitem[\protect\citeauthoryear{Zhao, Kitaura, Chuang, Prada, Yepes  \&
  Tao}{Zhao et~al.}{2015}]{Zhao:2015jga}
Zhao C.,  Kitaura F.-S.,  Chuang C.-H.,  Prada F.,  Yepes G.,   Tao C.,  2015,
  \mn@doi [Mon. Not. Roy. Astron. Soc.] {10.1093/mnras/stv1262}, 451, 4266

\makeatother
\end{thebibliography}

%%%%%%%%%%%%%%%%%%%%%%%%%%%%%%%%%%%%%%%%%%%%%%%%
%							Appendix										%
%%%%%%%%%%%%%%%%%%%%%%%%%%%%%%%%%%%%%%%%%%%%%%%%
\appendix

\section{Sample definition matching halo mass or halo abundance}\label{appendix1}

Here we show the difference in the results for samples defined by cutting at the same halo mass threshold as opposed to matching the abundance of N-body halos, as done in the main body of the article. We do this only for those methods that produce halo samples independently from the N-body (``predictive'') as the remaining methods work by reproducing abundance and mass distribution of the N-body samples, so selecting by mass thresholds or abundance matching is equivalent. In Table~\ref{samples_mass} we list the minimum halo mass and the average number of halos for these samples, for \icecola, \pinocchio and \peakpatch.

\begin{figure}
\begin{center}
\includegraphics[scale=0.98]{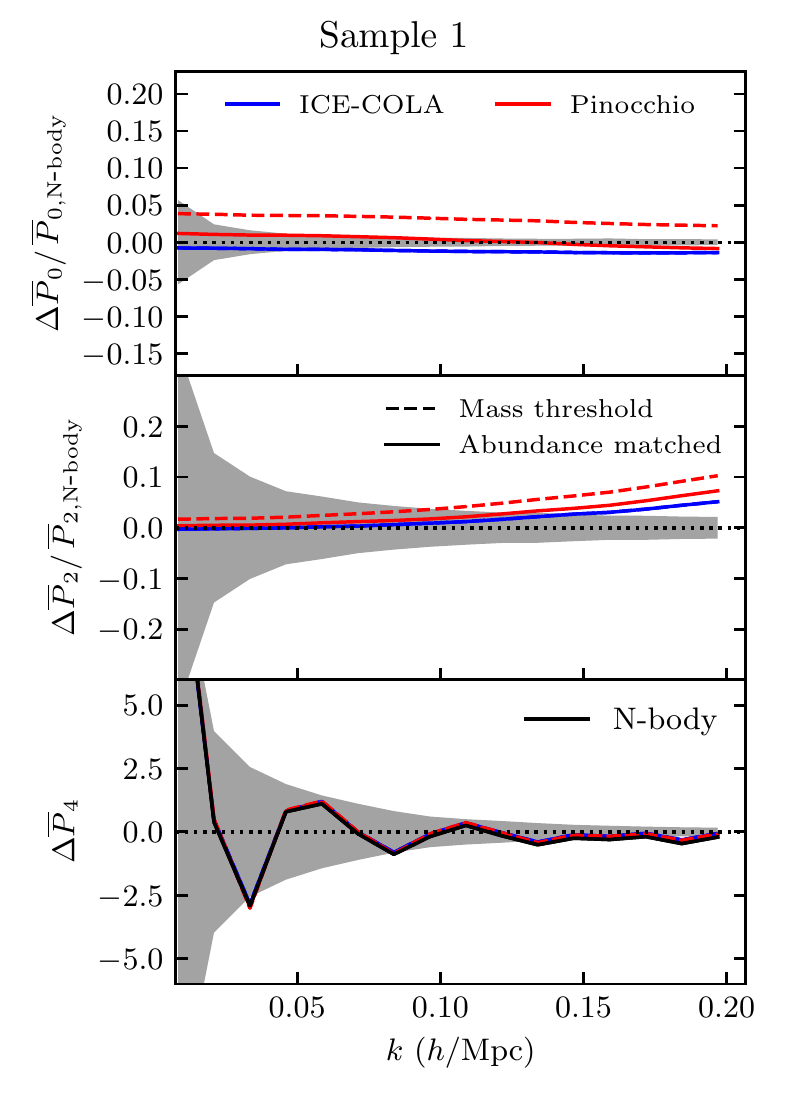}\\
\includegraphics[scale=0.98]{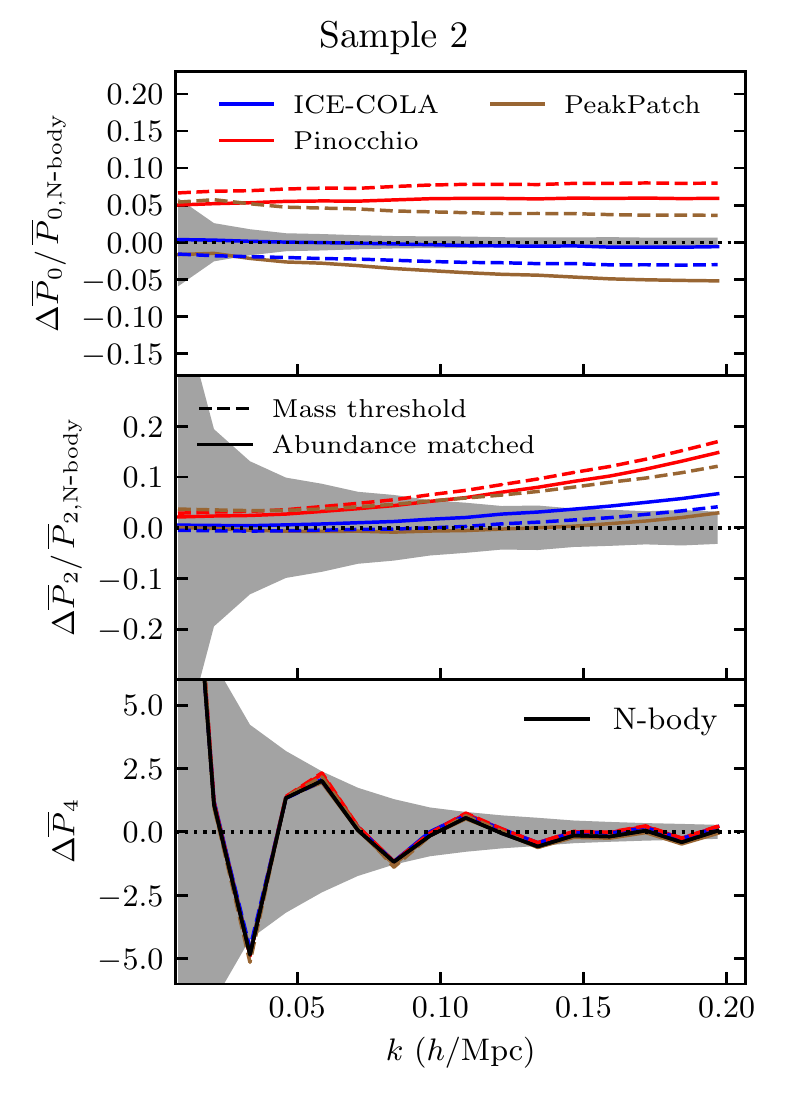}
\caption{Relative difference of the power spectrum monopole and quadrupole (hexadecapole) with respect to N-body (model) for the sample 1 (top plot) and sample 2 (bottom plot). Each plot shows the monopole (top panel), quadrupole (middle panel) and hexadecapole (bottom panel) for mass samples (dashed lines) and density samples (solid lines).}
\label{pkmlim_mass}
\end{center}
\end{figure}

\begin{figure}
\begin{center}
\includegraphics[scale=0.98]{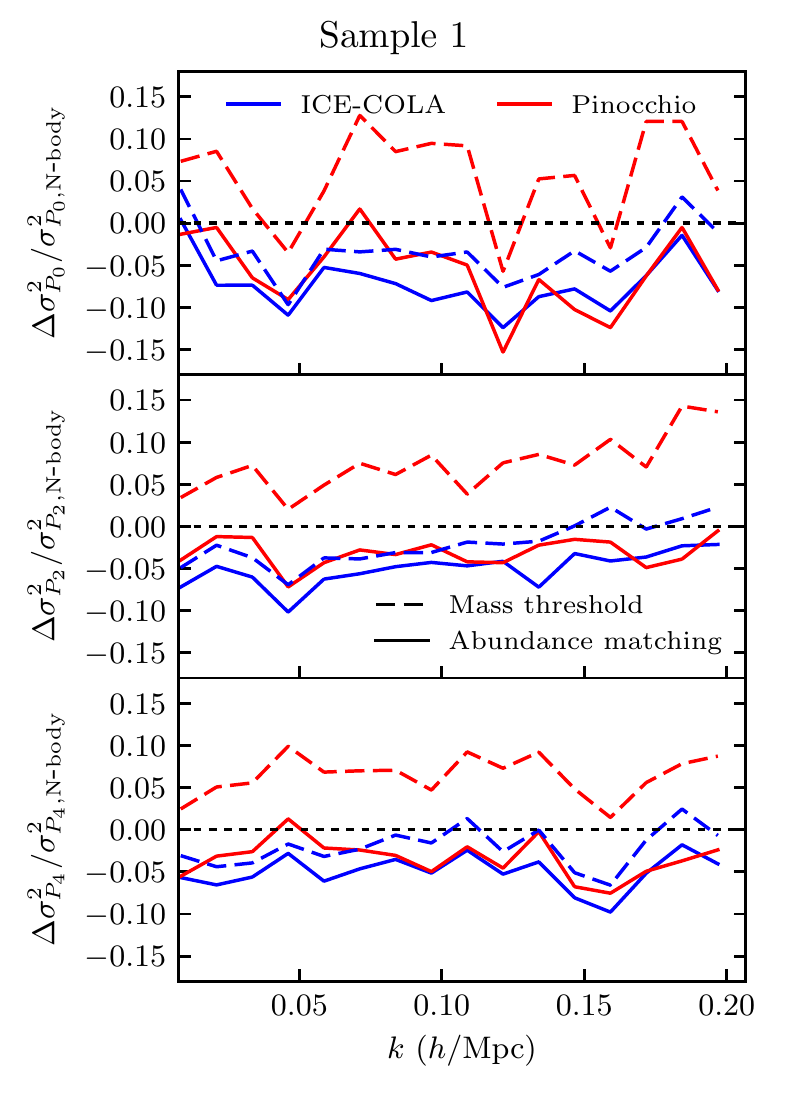}\\
\includegraphics[scale=0.98]{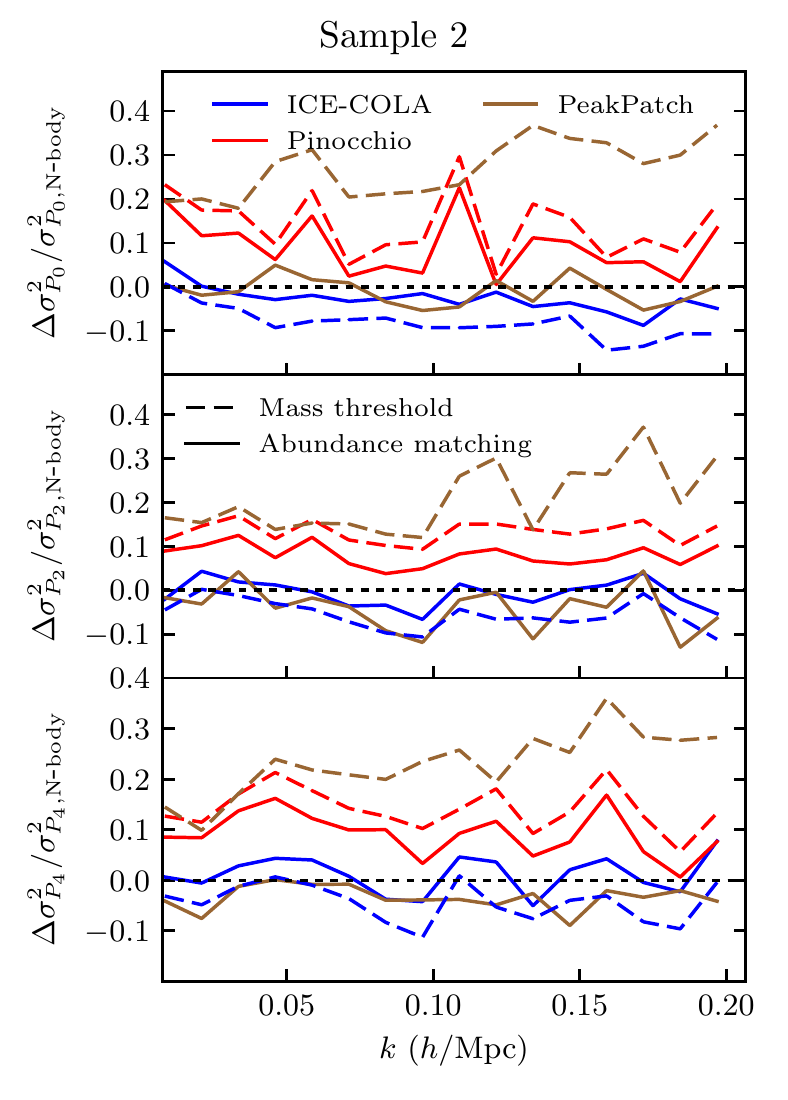}
\caption{Relative difference of the variance of power spectrum multipoles with respect to N-body for the sample 1 (top plot) and sample 2 (bottom plot). Each plot shows monopole (top panel), quadrupole (middle panel) and hexadecapole (bottom panel) for mass samples (dashed lines) and density samples (solid lines).}
\label{varmlim_mass}
\end{center}
\end{figure}

\begin{figure}
\begin{center}
\includegraphics{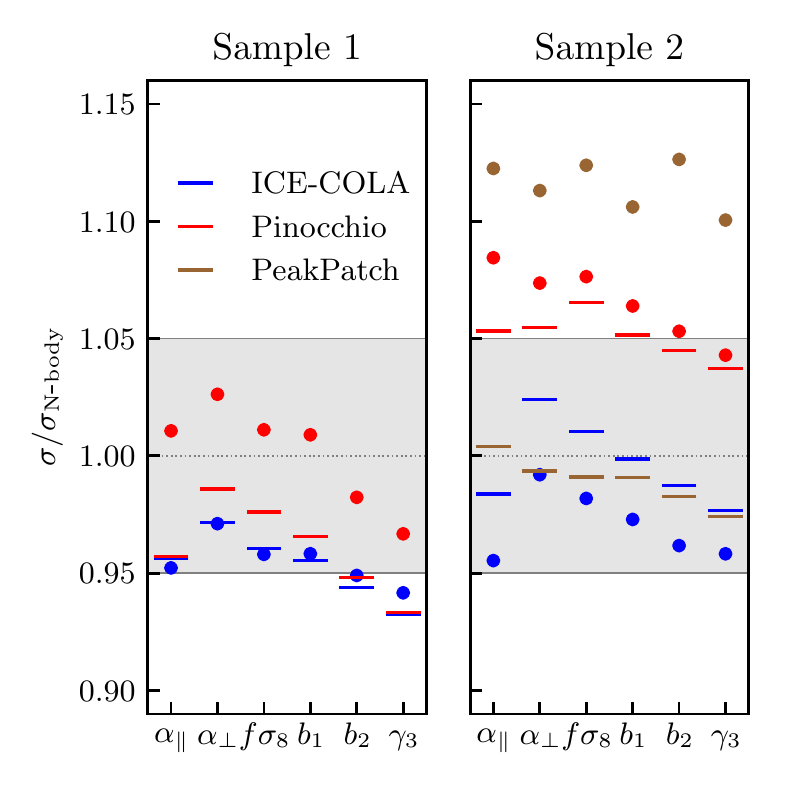}
\caption{Ratio of the error on cosmological and nuisance parameters with respect to N-body for sample 1 (left) and sample 2 (right) for mass samples (dots) and density samples (lines). The grey shaded area indicate $5\%$ differences.}
\label{err_ratio_mass}
\end{center}
\end{figure}

In Figures~\ref{pkmlim_mass}-\ref{varmlim_mass} we show the
average power spectrum multipoles and their variance respectively. In Fig. \ref{err_ratio_mass} we show how these differences translate into parameter errors, depicting with bars the results from abundance matching (as shown previously in Fig.~\ref{err_ratio}) and with dots the ones using mass thresholds.

In the case of \icecola the abundance matching works noticeably better in the most massive sample, while it is only marginally worse than the direct selection by mass at the low mass sample. This is compatible with the findings of \cite{IzardCrocceFosalba2016}, where the following trends were identified: i) halo masses are underestimated by $\sim 2\%-3\%$, and ii) at low masses the halo catalogs are incomplete (and the size of this effect will depend for example on the PM grid-size or the number of time-steps). At high masses (Sample 2) one can simply correct halo masses by doing abundance matching (or clustering matching). This effectively selects more massive halos, thus increasing the clustering amplitude and bringing the shot-noise level closer to its right value. Otherwise shot-noise is underestimated and so are the errors on parameters (since in this regime shot-noise is not negligible in the covariance). This is the behaviour seen in the right panel of Fig.~\ref{err_ratio_mass}. In turn, closer to the halo mass resolution of the catalogues (Sample 1) the effect is the opposite. Selecting halos by the same mass threshold as the N-body already shows slightly less clustering and abundance than the N-body. Imposing abundance matching shifts the mass scale and lowers the clustering amplitude even more (see top panel of Fig A1). Nonetheless in this mass regime the effect is small and the impact on parameter error is negligible. Both mass and abundance matching yield equivalent results in parameter errors for Sample 1, see left panel of Fig.~\ref{err_ratio_mass}.

For \pinocchio doing abundance matching always implies selecting effectively less massive halos and lowering the clustering amplitude. Since the raw clustering amplitudes are over-predicted at both Sample 1 and 2, the abundance matched helps in both cases. The performance of the abundance matched selection is better in Sample 2 only in terms of parameter errors, see Fig.~\ref{err_ratio_mass}. 

\peakpatch shows a similar behaviour in Sample 2 but this is much more expected. Halos in \peakpatch resemble by construction a more spherical structure closer perhaps to spherical overdensity samples, it is not calibrated to reproduce a FoF mass function. Hence, in this case, mass thresholds yield lower abundance than the N-body subFind / FoF halos, and hence higher clustering. The low abundance translates into higher shot-noise, and very over-estimated error bars (see second panel of Fig A3). This is solved by abundance matching which brings $\sigma$ and $\sigma_{\rm N-body}$ to within few percent.

Overall we find that abundance matching improves the agreement between samples in the approximate methods and the ones in N-body, at least in terms of parameter error bars. In article I of this series \citep{paper1} this comparison is extended to include samples defined by matching clustering amplitudes. 

\begin{table}
\begin{center}
\begin{tabular}{l|l|c|c}
Method & Selection Type & $M_{min}$ & ${\bar n}_{halos}$ \\
       &           &  ($\Msun$)  &  ($h^{3}$Mpc$^{-3}$)                  \\  
\hline
\multicolumn{3}{c}{Sample 1} \\
\hline 
N-body & mass threshold & $1.121 \times 10^{13}$ & $2.130\times 10^{-4}$ \\
\icecola & mass threshold & $1.121 \times 10^{13}$ & $2.057\times 10^{-4}$ \\
\pinocchio & mass threshold & $1.121 \times 10^{13}$ & $1.948\times 10^{-4}$ \\
\icecola & abund. match$^*$ & $1.086 \times 10^{13}$ & $2.122\times 10^{-4}$ \\
\pinocchio & abund. match$^*$ & $1.044 \times 10^{13}$ & $2.147\times 10^{-4}$ \\
\hline
\multicolumn{3}{c}{Sample 2} \\
\hline 
N-body & mass threshold & $2.670 \times 10^{13}$ & $5.441\times 10^{-5}$ \\
\icecola & mass threshold & $2.670 \times 10^{13}$ & $5.719\times 10^{-5}$ \\
\pinocchio & mass threshold & $2.670 \times 10^{13}$ & $5.258\times 10^{-5}$ \\
\peakpatch & mass threshold & $2.670 \times 10^{13}$ & $4.450\times 10^{-5}$ \\
\icecola &  abund. match$^*$ & $2.766 \times 10^{13}$ & $5.455\times 10^{-5}$ \\
\pinocchio & abund. match$^*$ & $2.631 \times 10^{13}$ & $5.478\times 10^{-5}$ \\
\peakpatch &  abund. match$^*$ & $2.355 \times 10^{13}$ & $5.439\times 10^{-5}$ \\
\end{tabular}
\end{center}
\caption{Characteristics of the mass threshold samples used in Appendix \ref{appendix1}, together with the abundance matched ones (marked with $^*$) used in the main body of the article.}
\label{samples_mass}
\end{table}%

\section{Errors associated with variance cancellation}\label{appendix_vc}

In this Appendix we estimate the error in our results due to the finite number of N-body simulations that we are using. To do this we make use of a larger set of 10,000 \pinocchio realisations with the same simulation and cosmological parameters.

We expect our results to depend to some extent on the number of the N-body realisations we are using as benchmark for the approximate methods \citep{HartlapSimonSchneider2007,TaylorJoachimiKitching2013,DodelsonSchneider2013,PercivalEtal2014}. The central figure in our comparison is the ratio between the parameter uncertainties obtained from the approximate mocks and from the N-body, so in this appendix we will attempt to estimate the error on such quantity as well as the advantages provided by matching the seeds of the 300 realisations. We will also consider more extensively how our results vary as a function of the largest wavenumber included in the analysis, $k_{max}$. 

In order to do this we will consider a simpler model depending on just two parameters that allows a fully analytical analysis of the likelihood function. We write the halo power spectrum as 
\beq
P_{model}(k)   = b_1 ^2\, P_L(k) +P_{SN}\,,
\eeq
where $b_1$ is the linear bias parameter and $P_{SN}$ is a constant accounting for the shot-noise component, while we assume the linear matter power spectrum $P_L(k)$ to be known. The likelihood function is therefore given by
\beq
\ln \mathcal{L}_{P}=-\frac12 \sum_{ij}^{k_{max}}\delta P(k_i)\,\left[C\right]^{-1}_{ij}\,\delta P(k_j)\,,
\label{eqlnL}
\eeq
where $\delta P\equiv P_{data}-P_{model}$ while $C_{ij}$ is the covariance matrix. Such model can be rewritten as
\beq
P_{model}   =  \sum_{\alpha=1}^2 p_\alpha\, \mathcal{P}_\alpha
\eeq
where $\left\{p_{\alpha}\right\}=\left\{ b_1^2, P_{SN} \right\}$
and $\left\{\mathcal{P}_\alpha\right\}  = \left\{ P_L(k) , 1\right\}$. Adding $p_0=-1$ and $\mathcal{P}_0=P_{data}$ we can also write
\beq
-\delta P=P_{model}-P_{data}   =  \sum_{\alpha=0}^2 p_\alpha\, \mathcal{P}_\alpha
\eeq
and therefore it is easy to see that we can rewrite the likelihood as
\beq
\ln \mathcal{L}_{P}  =  
-\frac12  \sum_{\alpha,\beta}^{2}\, p_\alpha\,p_\beta\,\mathcal{D}_{\alpha\beta}\,,
\eeq
where
\beq
\mathcal{D}_{\alpha\beta}\equiv \sum_{i,j}^{k_{max}}\,\mathcal{P}_{\alpha}(k_i)\,\left[C\right]^{-1}_{ij}\,\mathcal{P}_{\beta}(k_j)\,.
\eeq
In this way $\mathcal{L}_{P}$ is explicitly written as an exact, multivariate Gaussian distribution in the parameters $p_\alpha$. Clearly, once the quantities $\mathcal{D}_{\alpha\beta}$ are computed, we can evaluate any marginalisation analytically. This allows us to easily derive results for any value of $k_{max}$ and present them as a function of such quantity. 

\begin{figure}
\begin{center}
\includegraphics[width=0.45\textwidth]{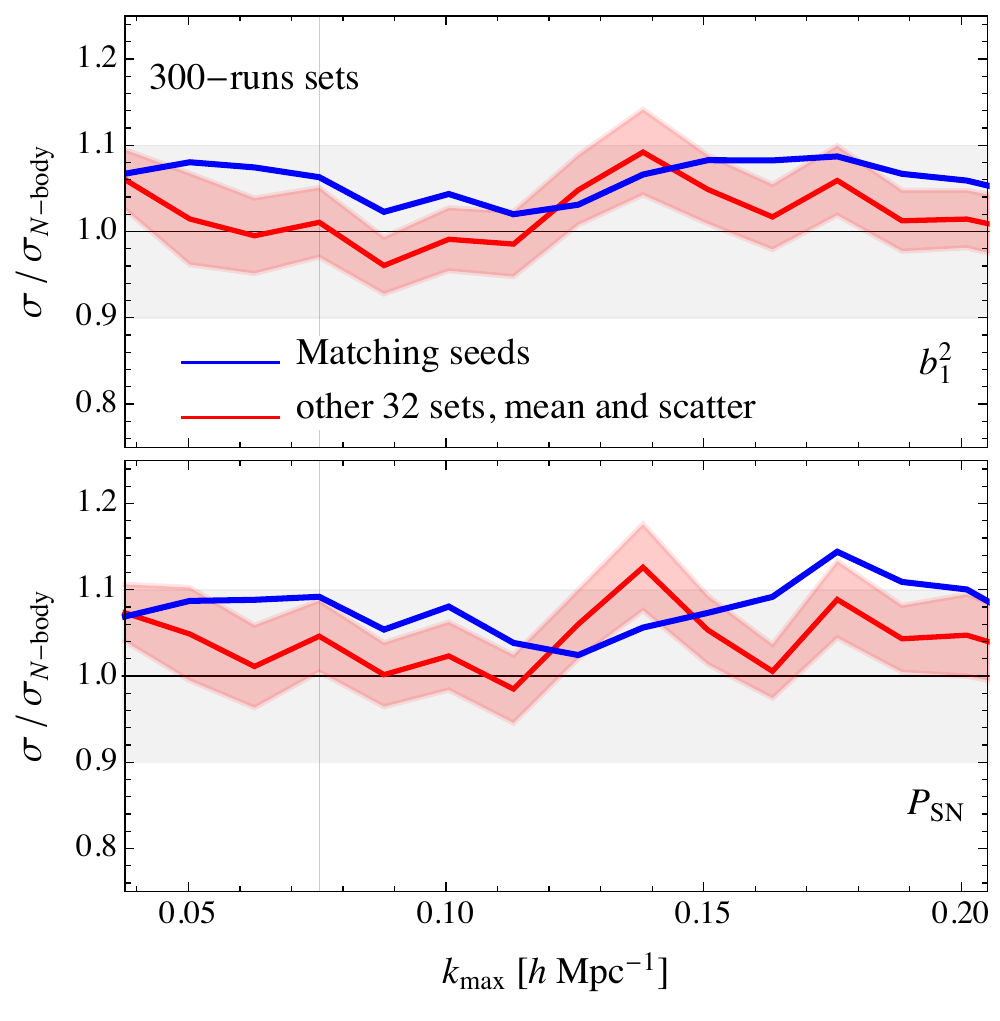}
\caption{Comparison between the ratio of marginalised uncertainties $\sigma/\sigma_{\rm N-body}$ obtained from the 300 \pinocchio and N-body realisations with matching seeds (blue curve) and the same quantity obtained from 32 distinct sets of 300 \pinocchio realisations and the same 300 N-body runs (red curve). The shaded area corresponds to the scatter across the 32 sets. The power spectrum model here depends on only two parameters, $b_1^2$ and $P_{SN}$. See the text for an explanation of the assumed likelihood function and power spectrum model. }
\label{fig:test33}
\end{center}
\end{figure}

As a first test, we compare the ratio of marginalised uncertainties $\sigma/\sigma_{\rm N-body}$ for 300 \pinocchio runs with matching seeds to the same quantity derived from 32 distinct sets of 300 \pinocchio runs with independent initial conditions. We should notice that this will not provide a full estimate of the error on the ratio $\sigma/\sigma_{\rm N-body}$ since the 32 sets are compared always with the same, single set of N-body realisations. Still, we can na\"ively expect the actual relative error on $\sigma/\sigma_{\rm N-body}$ to be $\sqrt{2}$ larger than the one we derive in this way. The results are shown in figure~\ref{fig:test33}. The blue curve shows the result for matching seeds as a function of $k_{max}$, top and bottom panel corresponding to the ratio of the marginalised uncertainties on, respectively, $b_1^2$ and $P_{SN}$. The red curve shows the mean across the 32 estimates of uncertainties ratio with non-matching seeds, while the shaded area shows its scatter. 

In the first place, we notice how the ratio can vary significantly as a function of $k_{max}$, even when we assume matching initial conditions. One should bear this in mind when discussing the recovered values of $\sigma/\sigma_{\rm N-body}$ presented in the main text. For non matching seeds the scatter across different values of $k_{max}$ is larger, as we can expect. Nonetheless in both cases the scatter is within the overall uncertainty on the ratio which we estimate to be $\sim 4\%-5\%$ (see below). 
 
\begin{figure}
\begin{center}
\includegraphics[width=0.45\textwidth]{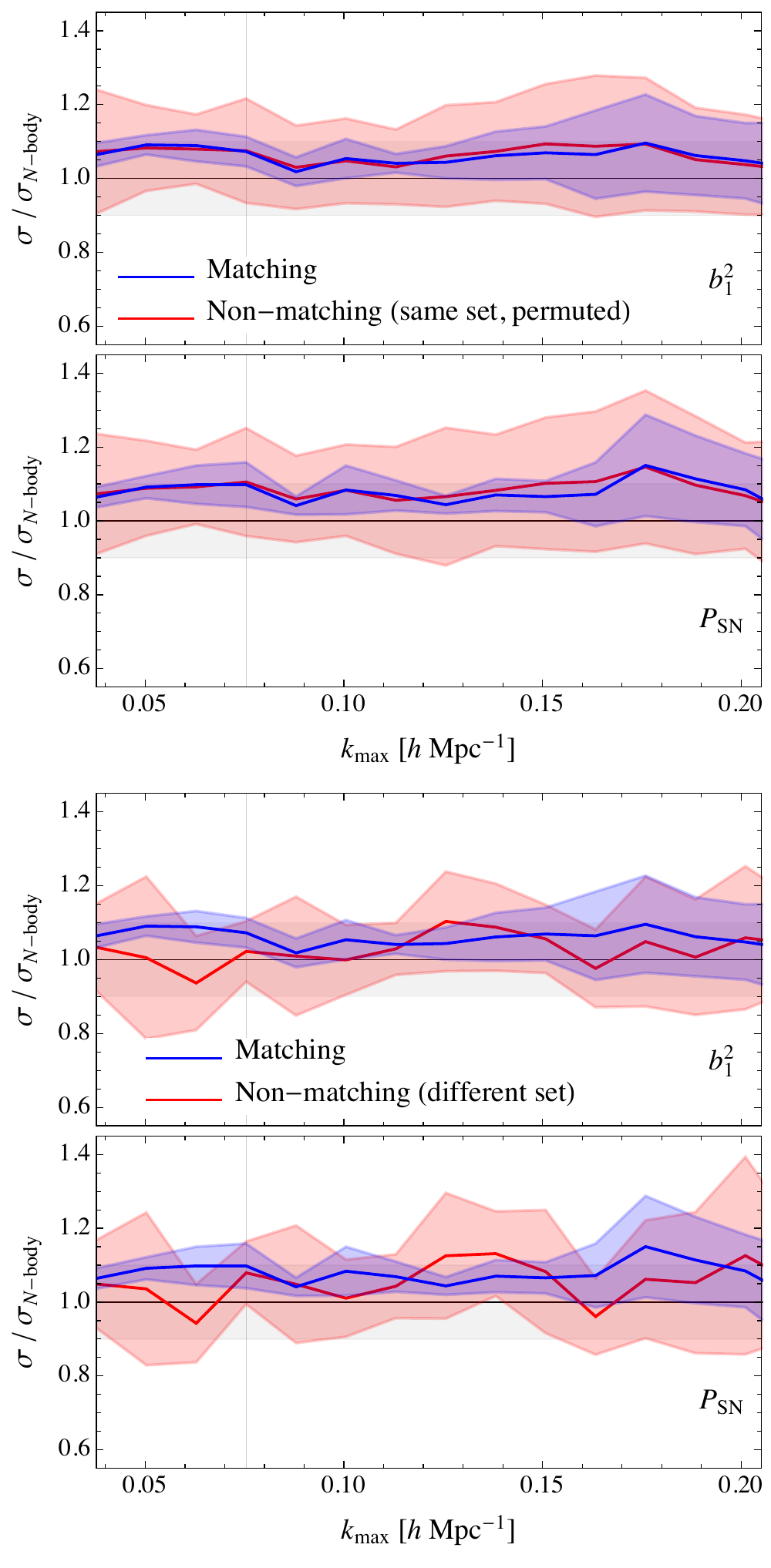}
\caption{The top panels show the mean and standard deviation (shaded area) of the ratios of the marginalised errors on the two parameters $b_1^2$ and $P_{SN}$ obtained from the \pinocchio covariance to the same errors obtained from the Minerva covariance. The blue line corresponds to the case of 6 sets of 50 realisations where we match the seed of the two methods. The red line to is obtained from the same sets but permuted in order to illustrate the case of non-matching seeds. The bottom panels show the same results for the matching seeds case, while the other case considers an independent set of \pinocchio runs. }
\label{fig:test50}
\end{center}
\end{figure}

The comparison of the mean over the independent sets and the matching seeds results shows in some cases quite large differences, but we should keep in mind that the former assumes always the same N-body set and therefore can be affected by large fluctuations in these realisations. Such fluctuations, on the other hand can be compensated in the matching seed case. From this figure we see that the scatter in the recovered value of $\sigma$ is $\sim 4\%-5\%$.
Remarkably, a direct application of the formulae in \cite{TaylorJoachimiKitching2013} (their Eqs. 55 \& 56) results in an estimate for this error, assuming $N_S=300$ runs and $N_D=3 \times 16$ data-points, of $\sim 4.5\%$, which is in very good agreement. Lastly, if we estimate the full error on the ratio to be $\sqrt{2}$ times the error estimated here from the scatter in the numerator of the ratio alone we recover a $7\%-8\%$ uncertainty. Differences between solid blue and red lines in Fig.~\ref{fig:test33} are well within these values.

As an additional test we subdivide the 300 mocks in 6 sets of 50 realisations. This number should ensure a marginally reliable estimation of the power spectrum covariance matrix, at least for small values of $k_{max}$. In this way we can evaluate the mean and the scatter of the ratio $\sigma/\sigma_{\rm N-body}$ between \pinocchio and the N-body results across 6 pairs and these can be computed for six matching pairs as well as 6 non-matching pairs obtained by permutations of the subsets. The results are shown in the upper panels of fig.~\ref{fig:test50}. This exercise shows that matching the initial conditions reduces the scatter on the uncertainties ratio by roughly a factor of two at large scales ($k\lesssim 0.1\kvecMpc$) with respect to the case of independent runs. Clearly such reduction is less significant at smaller scales, as one can expect. In the lower panels of fig.~\ref{fig:test50}, for the comparison among non-matching realisations shown by the red curve, we replace the six subsets of Pinocchio runs with other six, independent runs in order to avoid the permutation. Clearly the blue curve showing the mean ratio from matching seeds is the same as before. It may appear that in this more general comparison between non-matching realisations the mean value of the independent set is closer to one, indicating, apparently, a better performance of Pinocchio w.r.t. our main results. In fact this is simply reflecting an overall fluctuation of the first 300 N-body runs. Such fluctuation is captured to a large extent by the Pinocchio realisations with matching seeds allowing to trust the few percent overestimate of the parameters uncertainties provided by the approximate method.

To summarise, we estimate an uncertainty of about $7\%-8\%$ coming from the fact that our comparisons are limited to 300 N-body and approximate methods runs which is, however, reduced to about $4\%-5\%$ from adopting the same initial conditions in both cases.

\end{document}